\documentclass[preprint]{revtex4}
\usepackage{graphicx}

\begin{document}
\title{Mirror optical activity: efficient chiral sensing from electromagnetic parity indefiniteness}
\author{A. Ciattoni$^1$}
\email{alessandro.ciattoni@spin.cnr.it}
\affiliation{$^1$CNR-SPIN, c/o Dip.to di Scienze Fisiche e Chimiche, Via Vetoio, 67100 Coppito (L'Aquila), Italy}
\date{\today}
\begin{abstract}
Mirror symmetry is among the most fundamental concepts of physics and its spontaneous breaking at the molecular level allows chiral molecules to exist in two enantiomers that are mirror images of each other. The majority of chiro-optical effects routinely used to detect enantiomers in mixtures, as circular dichroism, relies on chiral sensitivity to photon circular polarization, thus not harnessing the full potentials of mirror symmetry breaking which also involves the radiation spatial profile. Here we show that the parity indefiniteness of the electromagnetic field interacting with chiral matter supports mirror optical activity, a novel chiro-optical effect where a chiral film, once probed by the mirror symmetric field of a nanoemitter, produces a near field whose spatial profile has broken mirror symmetry. The detection of near field dissymmetry provides an highly efficient chiral sensing technique, thus opening novel avenues to devising nanonophotonic schemes for ultra-efficient chiral discrimination of picogram quantities of molecules. We specialize the technique to nano-films with infrared chirality by using a swift electron in aloof configuration as the nanoemitter and an off-axis transparent conductor nanoparticle as the near field probe; the spatial dissymmetry factor of nanoparticle cathodoluminescence is one order of magnitude larger than circular dichroism, which is further enhanced to two orders if an additional graphene sheet is deposited on the film interface. 
\end{abstract}

\maketitle
Explicit symmetry breaking, the artificial lowering of the symmetry group of a system, is an ubiquitous and key physical concept since it creates phenomena characterized by intriguing asymmetric behavior, as first clarified by Pierre Curie in 1894 \cite{Curie}. For example, if time reversal is broken, photonic crystals with Dirac points display topological properties \cite{Halda,Ozawa} and they support topologically protected photonic edge modes with unidirectional propagation  \cite{Wangg}. Besides, the interplay of parity and time-reversal breaking in PT-symmetric photonic systems \cite{Fengg} underpins unidirectional reflectionless propagation in periodic structures \cite{Linnn} and angular momentum vortex lasing in microrings \cite{Miaoo}. Chiral phenomena involve only parity breaking and are of fundamental interest in physics with important implications for biology and life sciences. Molecular chirality, microscopically resulting from spontaneous parity breaking, explicitly breaks the mirror symmetry at the macroscopic scale and, accordingly, even homogeneous and isotropic chiral matter displays optical activity \cite{Barro}, its asymmetric interaction with left and right circularly polarized photons. Such asymmetry enables chiro-sensitive molecular detection through circular dichroism (the differential absorption of left and right circularly polarized light) or optical rotation (rotation of the polarization plane of light produced by the chiral medium) which is highly relevant in biological chemistry and pharmacology since many organic molecules are chiral. However, the mismatch between molecular size and radiation wavelength severely limits the strength of chiro-optical phenomena and accordingly a number of strategies have been proposed to achieve enhanced chiral sensing \cite{Colli,Munnn,Muuuu}. Paradigmatic examples are plasmon enhancement circular dichroism \cite{Govor,Abdul,Liuuu,Neste}, which exploits near field effects to reducing the molecular-field spatial scale mismatch, or superchiral fields, where the local enhancement of optical chirality \cite{Lipki,Vazqu} amplifies the differential circular dichroism signal \cite{Tang1,Tang2,Hend1}, both in plasmonic \cite{Hend2,Schaf,Petro,Staub} and dielectric setups \cite{Hoooo,Yaooo,Moham,Garci}. Other ingenious chiral sensing techiques has also been conceived which resort to femtosecond spettroscopy \cite{Rheee}, photoionization \cite{Janss,Beaul} and signal-reversing cavity ringdown polarimetry \cite{Sofik}.

Nearly all the chiro-optical phenomena hiterto considered rely on the effect of chiral matter on the polarization of light, as in optical rotation, or on the comparison of the medium response to two excitations of opposite polarization handedness, as in circular dichroism. However, the geometric operation of spatial reflection, in addition to change the direction of the electric (vector) and magnetic (pseudovector) fields, also involves the specular inversion of their profiles and the question arises as whether explicit mirror symmetry breaking is able to create chiro-optical effects displaying asymmetric spatial features. Here, we affirmatively answer this question by theoretically demonstrating that a mirror symmetric field (i.e. spatially coinciding with its mirror image) is reflected and transmitted by a chiral film into fields with indefinite parity, a phenomenon we refer to as mirror optical activity (MOA).  The spatial asymmetry of the reflected and transmitted fields is a consequence of the fundamental electromagnetic parity indefiniteness in chiral media and it is genuinely a novel chiro-optical effect since two geometrically identical samples, with opposite chirality (enantiomers) and excited by the same probe, produce fields which are mirror images of each other. Since any chiro-optical effect is magnified by near field interactions, MOA induced by a nanoemitter is particularly marked and the detection of the near field asymmetry provides an efficient technique for enantiometric discrimination. We specialize our general reasoning to a chiral molecular layer, with the possible inclusion of a graphene sheet at its interface, interacting with a fast electron in aloof configuration and we probe MOA through a transparent semiconductor nanoparticle whose position-dependent cathodoluminescence emission provides evidence of the near field asymmetry and chiral sensing.

\vfill \noindent {\bf ELECTROMAGNETIC PARITY INDEFINITENESS}

The starting point of our analysis is reflection invariance stating that when a complete experiment is subjected to mirror reflection, the resulting experiment should, in principle, be realizable (when the weak interaction can be neglected). In order to deal with an electromagnetic field ${\bf E}_\omega$,${\bf H}_\omega$ (with suppressed $e^{-i \omega t}$ factor) existing in a homegeneous and isotropic chiral medium, we note that the geometric reflection through a plane ${\bf{r}}' = {\mathcal{R}} {\bf{r}}$ (where $\mathcal{R}$ is the dyadic reversing the sign of the vector component normal to the plane) maps the field into its mirror image
\begin{eqnarray}
   {\bf{E}}'_\omega  \left( {\bf{r}} \right) &=& {\mathcal{R}}{\bf{E}}_\omega  \left( {{\mathcal{R}}{\bf{r}}} \right), \nonumber  \\
   {\bf{H}}'_\omega  \left( {\bf{r}} \right) &=& - {\mathcal{R}}{\bf{H}}_\omega  \left( {{\mathcal{R}}{\bf{r}}} \right), 
\end{eqnarray}
due to the vector and pseudovector nature of the electric and magnetic fields, whereas it maps the medium into an identical one but with opposite chirality (opposite enantiomeric medium) since microscopically each chiral molecule is reflected into its opposite enantiomer (see Fig.1a). Reflection invariance states that ${\bf E}'_\omega$,${\bf H}'_\omega$ is a physically admissible field in the opposite enantiomeric medium. In other words, if the field ${\bf E}_\omega$,${\bf H}_\omega$ satisfies Maxwell equations with constitutive relations
\begin{eqnarray} \label{chi_const}
 {\bf{D}}_\omega   &=& \varepsilon _0 \varepsilon {\bf{E}}_\omega   - \frac{i}{c}\kappa {\bf{H}}_\omega, \nonumber   \\ 
 {\bf{B}}_\omega   &=& \mu _0 {\bf{H}}_\omega   + \frac{i}{c}\kappa {\bf{E}}_\omega,
\end{eqnarray}
where $\varepsilon$ is the dielectric permittivity and $\kappa$ is the Pasteur parameter accounting for the electro-magnetic coupling produced by molecular chirality, the mirror image field ${\bf E}'_\omega$,${\bf H}'_\omega$ satisfies the same Maxwell equations but with $\varepsilon' = \varepsilon$ and $\kappa' = - \kappa$, as sketched in Fig.1a. This well known fact has a direct consequence which seems to be overlooked in the subject of chiro-optics: a mirror symmetric field, i.e. a field coinciding with its mirror image ${\bf E}_\omega = {\bf E}'_\omega$, ${\bf H}_\omega ={\bf H}'_\omega$, can not exist in a chiral medium since it should be an admissible field also in the opposite enantiomeric medium (see Fig.2a).  Evidently, the argument breaks down in the absence of chirality ($\kappa = 0$) since an achiral medium coincides with its mirror image. 

In order to thoroughly characterize the above dissymmetry, it is useful to focus on the electric field satisfying the equation
\begin{equation} \label{Maxwell}
- \nabla  \times \nabla  \times {\bf{E}}_\omega   + k_0^2 \left( {\varepsilon  - \kappa ^2 } \right){\bf{E}}_\omega   =  2k_0 \kappa \nabla  \times {\bf{E}}_\omega  
\end{equation}
where $k_0 = \omega / c$, which is fully equivalent to Maxwell equations and whose explicit broken mirror symmetry is evident since its LHS is a vector whereas its RHS is a pseudovector scaled by the chirality parameter $\kappa$. Consider now the symmetric and antisymmetric parts of the electric field given by
\begin{eqnarray} \label{SymAsym}
 {\bf{E}}_\omega ^{\rm S }  &=& \frac{1}{2}\left( {{\bf{E}}_\omega   + {\bf{E}}'_\omega  } \right), \nonumber  \\ 
 {\bf{E}}_\omega ^{\rm A }  &=& \frac{1}{2}\left( {{\bf{E}}_\omega   - {\bf{E}}'_\omega  } \right), 
\end{eqnarray}
which have definite parity since they are eigenstates of the reflection operator $\hat R$  acting on vector fields as $\hat R\left[ {{\bf{A}}\left( {\bf{r}} \right)} \right] = \mathcal{R}{\bf{A}}\left(  {\mathcal{R}{\bf{r}}} \right)$, with eigenvalues $+1$ and $-1$, respectively. As a consequence ${\bf{E}}_\omega$ is a mirror symmetric field (MSF) or a mirror antisymmetric field (MAF) if ${\bf{E}}_\omega ^{\rm A} =0$ or ${\bf{E}}_\omega ^{\rm S} =0$, respectively. Inserting the decomposition ${\bf{E}}_\omega   = {\bf{E}}_\omega ^{\rm S}  + {\bf{E}}_\omega ^{\rm A}$ into Eq.(\ref{Maxwell}) and separating the symmetric and antisymmetric parts of the resultant equation we obtain
\begin{eqnarray} \label{coupling}
- \nabla  \times \nabla  \times {\bf{E}}_\omega ^{\rm S}  + k_0^2 \left( {\varepsilon  - \kappa ^2 } \right){\bf{E}}_\omega ^{\rm S}  &=&   2k_0 \kappa \nabla  \times {\bf{E}}_\omega ^{\rm A}, \nonumber  \\ 
- \nabla  \times \nabla  \times {\bf{E}}_\omega ^{\rm A}  + k_0^2 \left( {\varepsilon  - \kappa ^2 } \right){\bf{E}}_\omega ^{\rm A}  &=&   2k_0 \kappa \nabla  \times {\bf{E}}_\omega ^{\rm S},
\end{eqnarray}
which vividly show how chirality couples the definite parity  parts of the field. What is essential for our purposes is that these equations make it clear that  ${\bf{E}}_\omega ^{\rm S}$ and ${\bf{E}}_\omega ^{\rm A}$ can not separately vanish (unless the extremely critical condition $\varepsilon = \kappa^2$ is achieved) thus showing that neither MSFs (as proven above) or MAFs can exist in a chiral medium. We conclude that the field interacting with chiral matter is always an indefinite parity field (IPF) and that this is the most comprehensive consequence of macroscopic explicit symmetry breaking produced by molecular chirality. It is worth emphasizing that in achiral media the symmetric and antysimmetric parts of the field are not coupled (see Eqs.(\ref{coupling}) with $\kappa = 0$) and consequently MSFs, MAFs and IPFs are all physically admissible fields.

\begin{figure*}[!] \label{Fig1}
\includegraphics[width=1\textwidth]{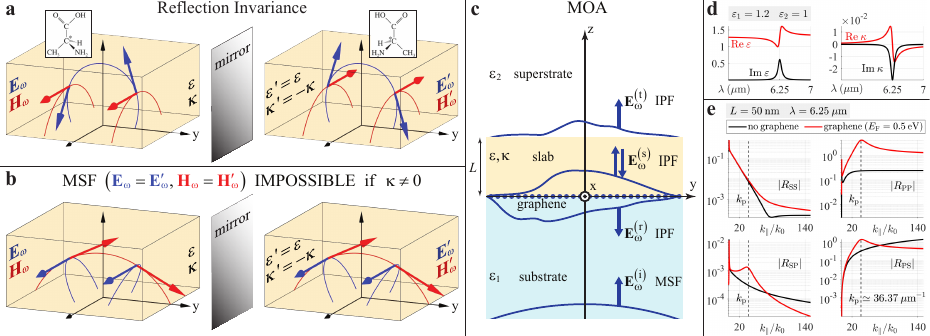}
\caption{{\bf Electromagnetic parity indefinitenss in chiral media and mirror optical activity (MOA)}. {\bf a}. If the field ${\bf E}_\omega$,${\bf H}_\omega$ exists in a homogeneous chiral medium, its mirror image ${\bf E}'_\omega$,${\bf H}'_\omega$ is a physically admissible field in the opposite enantiomeric medium. {\bf b}. A mirror symmetric field (MSF) can not exist in a chiral medium since it should exist in the opposite enantiomeric medium as well. {\bf c}. MOA produced by a chiral slab sandwitched between two achiral dielectrics with the potential inclusion of a graphene sheet. The incident (i) MSF triggers the excitation of the slab (s) indefinite parity field (IPF) in turn producing reflected (r) and transmitted (t) IPFs. {\bf d}. Parameters of the setup used for numerical evaluations. {\bf e}. Reflection coefficients of a $\rm 50 \: nm$ thick chiral slab at molecular resonance $\lambda = 6.25 \: \mu {\rm m}$ as functions of the (normalized) transverse photon momentum, with and without graphene inclusion.}
\end{figure*}

\vfill \noindent {\bf MIRROR OPTICAL ACTIVITY}

The peculiarity of electromagnetic parity indefiniteness rises the question of what happens when a chiral sample surrounded by an achiral environment is probed by an external field with definite parity. Since the sample only hosts IPFs, it turns out that the field it produces in the environment can not have definite parity thus showing dissymmetric features. Such asymmetric response to a symmetric stimulus is a genuinely novel chiro-optical effect, the mirror optical activity (MOA), since it is entirely due to sample chirality which, in turn, can be measured through the field dissymmetry detection.

To prove these observations and make them quantitative, we consider a chiral slab ($\varepsilon$, $\kappa$) surrounded by a substrate ($\varepsilon_1$) and a superstrate ($\varepsilon_2$) which are achiral transparent dielectrics, as sketched in Fig.1c. We also incorporate a graphene sheet lying at the slab-substrate interface since, although not essential for the onset of MOA, the excitation of plasmon polaritons combined with field dissymmetry triggers a novel near field interferometric mechanism able to enhance the enantiomeric sensing (see below). Due to its planar geometry and homogeneity, the setup is geometrically (but not physically) left invariant by reflections through any plane orthogonal to the interfaces so that, to isolate electromagnetic asymmetric features entirely due to chirality, we hereafter focus on reflections through the $xz$ plane, i.e. $\mathcal{R} =  {{\bf{e}}_x {\bf{e}}_x - {\bf{e}}_y {\bf{e}}_y  + {\bf{e}}_z {\bf{e}}_z } $. The slab is probed by an incident MSF, ${\bf{E}}_\omega ^{\left( \rm i \right)}   \left( {\bf{r}} \right) = \mathcal{R}{\bf{E}}_\omega ^{\left( \rm i \right)}  \left( {\mathcal{R}{\bf{r}}} \right)$, in the substrate which is given in full generality by
\begin{equation} \label{i-field}
{\bf{E}}_\omega ^{\left( \rm i \right)}  = \int {d^2 {\bf{k}}_\parallel  } e^{i{\bf{k}}_\parallel   \cdot {\bf{r}}_\parallel  }   e^{ik_{1z} z} \left[ {U_{\rm S}^{\left(\rm  i \right)} {\bf{u}}_{\rm S}  + U_{\rm P}^{\left(\rm  i \right)} \left( {{\bf{u}}_{\rm P}  - \frac{{k_\parallel  }}{{k_{1z} }}{\bf{e}}_z } \right)} \right],
\end{equation}
where the subscript $\parallel$ hereafter labels a vector parallel to the slab (${\bf{A}}_\parallel   = A_x {\bf{e}}_x  + A_y {\bf{e}}_y$), $k_{1z}  = \sqrt {k_0^2 \varepsilon _1  - k_\parallel ^2 }$, ${\bf{u}}_{\rm S}  = {\bf{e}}_z  \times {\bf{k}}_\parallel  /k_\parallel$,  ${\bf{u}}_{\rm P}  = {\bf{k}}_\parallel  /k_\parallel$ are the transverse electric (S) and transverse magnetic (P) unit vectors in momentum space and $U_{\rm S}^{\left( \rm i \right)}$, $U_{\rm P}^{\left( \rm i \right)}$ are the corresponding amplitudes. The definite and positive parity ($+1$) of the incident field amounts in momentum space to the antisymmetry and symmety of its S and P amplitudes, respectively, i.e. $U_{\rm S}^{\left( \rm i \right)} ( {{\bf{k}}_\parallel  } ) =  - U_{\rm S}^{\left( \rm i \right)} ( {{\mathcal{R}}{\bf{k}}_\parallel  } )$, $U_{\rm P}^{\left( \rm i \right)} ( {{\bf{k}}_\parallel  } ) =  U_{\rm P}^{\left( \rm i \right)} ( {{\mathcal{R}}{\bf{k}}_\parallel  } )$. An IPF ${\bf{E}}_\omega ^{\left( \rm s \right)}$ arises inside the slab together with reflected ${\bf{E}}_\omega ^{\left( \rm r \right)}$ and transmitted ${\bf{E}}_\omega ^{\left( \rm t \right)}$ fields in the substrate and superstrate, respectively (see Methods and Supporting Information). To show that the reflected field (and analogously the transmitted field) is an IPF we note that its symmetric and antisymmetric parts  are
\begin{eqnarray} \label{MOAr}
 {\bf{E}}_\omega ^{\left( \rm r \right){\rm S}}  &=& \int {d^2 {\bf{k}}_\parallel  } e^{i{\bf{k}}_\parallel   \cdot {\bf{r}}_\parallel  }  e^{ - ik_{1z} z} \left[ {R_{\rm SS} U_{\rm S}^{\left( \rm i \right)} {\bf{u}}_{\rm S}  + R_{\rm PP} U_{\rm P}^{\left( \rm i \right)} \left( {{\bf{u}}_{\rm P}  + \frac{{k_\parallel  }}{{k_{1z} }}{\bf{e}}_z } \right)} \right] , \nonumber  \\ 
 {\bf{E}}_\omega ^{\left( \rm r \right){\rm A}}  &=& n \int {d^2 {\bf{k}}_\parallel  } e^{i{\bf{k}}_\parallel   \cdot {\bf{r}}_\parallel  }  e^{ - ik_{1z} z} \left[ {R_{\rm SP} U_{\rm P}^{\left( \rm i \right)} {\bf{u}}_{\rm S}  + R_{\rm PS} U_{\rm S}^{\left( \rm i \right)} \left( {{\bf{u}}_{\rm P}  + \frac{{k_\parallel  }}{{k_{1z} }}{\bf{e}}_z } \right)} \right], 
\end{eqnarray}
where $n = \sqrt{\varepsilon \kappa^2}/\kappa$ and $R_{ij}$ is the reflection coefficient providing the contribution to the $i=\{ \rm S,P \}$ amplitude of the reflected field produced by the $j=\{ \rm S,P \}$ amplitude of the incident field (see Supporting Information). Now ${\bf{E}}_\omega ^{\left( \rm r \right){\rm S}}$ and ${\bf{E}}_\omega ^{\left( \rm r \right){\rm A}}$ vanish only if $U_{\rm S}^{\left( \rm i \right)}$ and $U_{\rm P}^{\left( \rm i \right)}$ simultaneously vanish, i.e. only if there is no incident field, thus proving that the reflected field is always an IPF. Remarkably ${\bf{E}}_\omega ^{\left( \rm r \right){\rm A}}$ only contains the mixing coefficients $R_{\rm SP}$ and $R_{\rm PS}$ which vanish in the absence of chirality ($\kappa = 0$) since in this case the $\rm S$ and $\rm P$ polarizations are independent. This shows that MOA, the emergence of the antisymmetric part of the field reflected by an homogeneous chiral slab probed by an incident MSF, is an effect entirely due to slab chirality. 

Another specific chiro-optical feature of MOA is that two identical slabs but filled with opposite enantiomeric media and probed by the same incident MSF produce reflected field which are mirror images of each other. Indeed, since the reflection coefficients $R_{ij}$ are left invariant by chirality reversal $\kappa \rightarrow - \kappa$ (see Supporting Information), the symmetric and antisymmetric parts of the field reflected by the slab with chirality $-\kappa$ are ${\bf{E}}_\omega ^{\left( \rm r \right){\rm S}}$ and $-{\bf{E}}_\omega ^{\left( \rm r \right){\rm A}}$, due to $n$ in the second of Eqs.(\ref{MOAr}), and these are precisely the symmetric and antisymmetric parts of ${ {{\bf{E}}_\omega ^{\left( \rm r \right)} } }'$, the mirror image of the field reflected by slab with chirality $\kappa$. Such MOA feature suggests viable enantiometric sensing techniques where the detection of the reflected field asymmetry enables the discrimination of different enantiomerically pure samples or the measurement of the enantiomeric excess of a mixture.

To discuss MOA more in depth, we focus on the reflection coefficients $R_{ij}$ pertaining a specific chiral slab of thickness $L = 50 \: {\rm nm}$ lying on a dielectric substrate in vacuum (the setup electromagnetic parameters are detailed in Fig.1d). In view of the potential graphene inclusion and its infrared effectiveness, we have chosen the chiral medium as a dielectric matrix with dispersed molecules of $\alpha$-alanine due to its marked chiral response resulting from a vibrational mode at $\rm 1600 \: cm^{-1}$ \cite{Tulio,Jahni}. In Fig.1e we plot the moduli of the reflection coefficients $|R_{ij}(k_\parallel)|$ at molecular resonance $\lambda = 6.25 \: \mu {\rm m}$ as functions of the transverse photon momentum (its direction not affecting the coefficients by isotropy, see Supporting Information) both without graphene (black curves) and with graphene of Fermi energy $E_{\rm F} = 0.5 \: eV$ (red curves). In the grafene-free case, the most striking feature is that the relevant MOA coefficients $R_{\rm SP}$ and $R_{\rm PS}$ (see the second of Eqs.(\ref{MOAr})) display quite opposite behaviors at large photon momenta $k_\parallel \gg k_0$, the first being small and fading, the second increasing and reaching  relatively large values with $|R_{\rm PS}| \gg |R_{\rm SP}|$. Analogous behaviors are shown by the coefficients $R_{\rm SS}$ and $R_{\rm PP}$ ruling the symmetric part of the reflected field.  Therefore MOA is enhanced if triggered by an incident MSF with deep sub-wavelength spatial features, as the field of a nano-emitter, whose photons are mainly $\rm S$ polarized since these traits highlight ${\bf{E}}_\omega ^{\left( \rm r \right){\rm A}}$ at the expense of ${\bf{E}}_\omega ^{\left( \rm r \right){\rm S}}$. The qualitative behavior of the reflection coefficients is not altered by graphene inclusion which basically yields the plasmon resonance peak at $k_{\rm p} \simeq 36.17 \: k_0 = 36.37 \: \mu {\rm m}^{-1}$ (except for the purely $\rm S$ coefficient $R_{\rm SS}$).

Before delving into the analysis of a feasible scheme for observing MOA, we elucidate its underpinning electromagnetic mechanism to explain the behavior of the mixing reflection coefficients in momentum space. The key fact is that photons are circularly polarized in chiral media so that their $\rm S$ and $\rm P$ components are not independent (see Methods and Supporting Information), the corresponding $\rm S$-$\rm P$ effective coupling having three main relevant consequences. First, it is responsible for the mixing reflection processes described by the coefficients $R_{\rm SP}$ and $R_{\rm PS}$ where incoming photons of a kind are reflected into photons of the other kind. Second, it makes such mixing reflection processes very sensitive also to the incoming magnetic field
\begin{equation}\label{i-Hfield}
{\bf{H}}_\omega ^{\left(\rm i \right)}  = \frac{1}{{Z_0 }} \int {d^2 {\bf{k}}_\parallel  } e^{i{\bf{k}}_\parallel   \cdot {\bf{r}}_\parallel  }  {e^{ik_{1z} z} \left[ {\frac{{k_0 \varepsilon _1 }}{{k_{1z} }}U_{\rm P}^{\left(\rm i \right)} {\bf{u}}_{\rm S}  - \frac{{k_{1z} }}{{k_0 }} U_{\rm S}^{\left(\rm i \right)} \left( {{\bf{u}}_{\rm P}  - \frac{{k_\parallel  }}{{k_{1z} }}{\bf{e}}_z } \right)} \right]},
\end{equation}
where $Z_0=\sqrt{\mu_0/\epsilon_0}$, due to the manifest $\rm S$-$\rm P$ role switch of its amplitudes. Third, it dramatically reduces the $\rm S$ component of large momentum photons in the chiral slab, since their $\rm P$ and $z$ components combine, by isotropy, in a nearly circularly polarized state for $k_\parallel \gg k_0$ (transverse spin of TM evanescent waves \cite{Fortu}) and the orthogonal S component is suppressed to preserve the photon circular polarization (see Methods and Supporting Infromation). Now the reflection process $\rm P \rightarrow S$ ($U_{\rm S}^{\left(\rm i \right)}=0$, $U_{\rm P}^{\left(\rm i \right)}\neq 0$), where an incoming photon of kind $\rm P$ is reflected into a photon of kind $\rm S$, is extremely inefficient for $k_\parallel \gg k_0$ since the magnetic field contribution is negligible (due to $k_{1z} \simeq ik_\parallel$ in the denominator) and the field in the slab has a negligible $\rm S$ component, thus explaining the asymptotic fall-off of $R_{\rm SP}$. Conversely, the reflection process $\rm S \rightarrow P$ ($U_{\rm S}^{\left(\rm i \right)}\neq 0$, $U_{\rm P}^{\left(\rm i \right)} = 0$) is boosted for $k_\parallel \gg k_0$ since the magnetic field contribution is very large (due to $k_{1z} \simeq ik_\parallel$ in the numerator) and field in the slab is almost of $\rm P$ kind, thus explaining the dramatic enhancement of $R_{\rm PS}$ in the large photon momentum regime.

\begin{figure*}[!] \label{Fig2}
\includegraphics[width=1\textwidth]{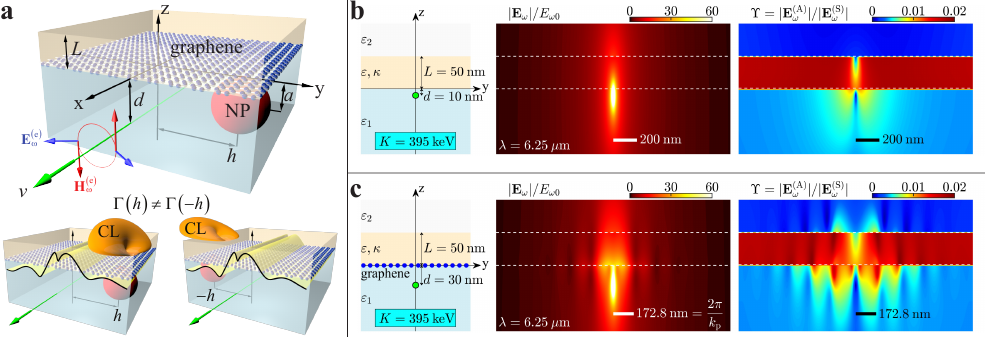}
\caption{{\bf MOA triggered by fast electrons}. {\bf a}. A fast electron traveling at velocity $v$ through the substrate in aloof configuration generates a MSF triggering MOA while the off-axis nanoparticle (NP) at $y = h$ locally probes the near field. Due to the asymmetric near field (yellow), the cathodoluminescence (CL) emission probabilities $\Gamma(h)$ and $\Gamma(-h)$ of the nanoparticle located at two mirror symmetrical positions are unequal, their difference providing MOA detection and chiral sensing. {\bf b,c}. Normalized modulus of the total field ${\bf E}_\omega$ and its dissymmetry factor $\Upsilon$ in two setups (without nanoparticle), ({\bf b}) without graphene and ({\bf c}) with graphene inclusion. $\Upsilon$ provides a local estimation of MOA efficiency.}
\end{figure*}

\vfill \noindent {\bf FAST ELECTRON SOURCE}

An electron traveling at relativistic speed is a very appropriate source to trigger MOA since the cylindrically symmetric field it creates is a MSF with a strong near field. In addition, electron velocity affects the inclination of the electric field vector thus providing management of its $\rm S$-$\rm P$ content in momentum space which is useful for MOA. For these reasons, we hereafter focus on the excitation scheme sketched in Fig.2a where an electron of charge $-e<0$ moves in the substrate at constant velocity $v>0$ along a trajectory parallel to the slab interface at distance $d$ (the nanoparticle NP will be included in the next section). The field generated by the electron \cite{deAba} is (for $\omega >0$)
\begin{eqnarray} \label{Ee}
 {\bf{E}}_\omega ^{\left( \rm e \right)}  &=& \frac{{E_{\omega 0} }}{{\varepsilon _1 \beta ^2 \gamma }}e^{i\frac{\omega }{v}x} \left[ { - K_1 \left( {\frac{{\omega R}}{{v\gamma }}} \right){\bf{e}}_R  + \frac{i}{\gamma }K_0 \left( {\frac{{\omega R}}{{v\gamma }}} \right){\bf{e}}_x } \right], \nonumber \\ 
 {\bf{H}}_\omega ^{\left( \rm e \right)}  &=& \frac{{E_{\omega 0} }}{{Z_0 \beta \gamma }}e^{i\frac{\omega }{v}x} \left[ { - K_1 \left( {\frac{{\omega R}}{{v\gamma }}} \right){\bf{e}}_\Phi  } \right],
\end{eqnarray}
where $E_{\omega 0}  = {eZ_0 \omega }/{(4\pi ^2 c)}$ is a field amplitude, $\beta = v/c$, $\gamma  = 1/\sqrt{1 - \varepsilon_1 \beta ^2}$ is the Lorentz contraction factor, $K_n$ are the modified Bessel function of the second kind and $(R,\Phi)$ are cylindrical coordinates coaxial with the charge trajectory with coordinate unit vectors ${\bf{e}}_R$ and ${\bf{e}}_\Phi$. The mirror symmetry of the electron field with respect to reflection $y \rightarrow -y$ is evident and its deep subwavelength features are particularly marked in the sub-Cherenkov regime $v < c/\sqrt \varepsilon_1$ we here consider, due to its radial exponential decay. 

In Fig.2b we consider an electron beam of kinetic energy $K = 395 \: {\rm keV}$ located at $d=10 \: {\rm nm}$ away from the $\rm 50 \: nm$ thick slab of the above considered setup (Figs.1d,e) and we plot the normalized modulus of the total field $|{\bf{E}}_\omega|/E_{\omega 0}$  at molecular resonance $\lambda = 6.25 \: \mu m$, together with its field dissymmetry factor
\begin{equation}
\Upsilon = \frac{|{\bf E}_\omega^{(\rm A)}|}{|{\bf E}_\omega^{(\rm S)}|}.
\end{equation} 
The electron field decay length is of the order of $\rm 200 \: nm$ and hence the chosen distance $d$ enables this field to penetrate through the slab and to experience molecular chirality. This is explicitly demonstrated by the plot of $\Upsilon$ revealing a field asymmetry percentage of the order $2\%$ within the slab and, most importantly for our purposes, of the order of $1\%$ in the substrate portion surrounding the electron beam. The non-vanishing of $\Upsilon$ outside the slab markedly signals the occurrence of MOA. In Fig.2c we consider the same situation of Fig.2b but with the inclusion of graphene and with the electron beam located at $d=30 \: {\rm nm}$ away from the slab. The total field profile now displays a modulation parallel to the slab of period $\rm 172.8 \: nm = 2 \pi / k_{\rm p}$ (see Fig.1e) resulting from the interference of the two excitated graphene plasmon polaritons with parallel momenta $k_y= k_p$ and $k_y= -k_p$. The impact of such counterpropagating plasmons on MOA efficiency is displayed in the plot of $\Upsilon$ which dramatically reveals a modulation and an overall enhancement of the field asymmetry percentage (compare with Fig.2b), a novel near field interferometric phenomenon resulting from the interplay of plasmon excitation and field asymmetry. Indeed, symmetry and antisymmetry in momentum space of ${\bf{E}}_\omega ^{\left( \rm r \right){\rm S}}$ and ${\bf{E}}_\omega ^{\left( \rm r \right){\rm A}}$ (see Eqs.(\ref{MOAr})), respectively, compel their leading plasmon contributions to have the $y$ depencies  ${\bf{E}}_\omega ^{\left(\rm r \right){\rm S}}  \sim   {\rm cos} \left( {k_p y} \right)$ and $ {\bf{E}}_\omega ^{\left(\rm r \right){\rm A}}  \sim {\rm sin} \left( {k_p y} \right)$ thus explaining the enhancement of $\Upsilon$ close to those points where $|{\bf{E}}_\omega ^{\left(\rm r \right){\rm S}}|$ and $|{\bf{E}}_\omega ^{\left(\rm r \right){\rm A}}|$ are respectively minimum and maximum at once, due to their $\pi/2$ dephasing (see Supporting Information).

\begin{figure*}[!] \label{Fig3}
\includegraphics[width=1\textwidth]{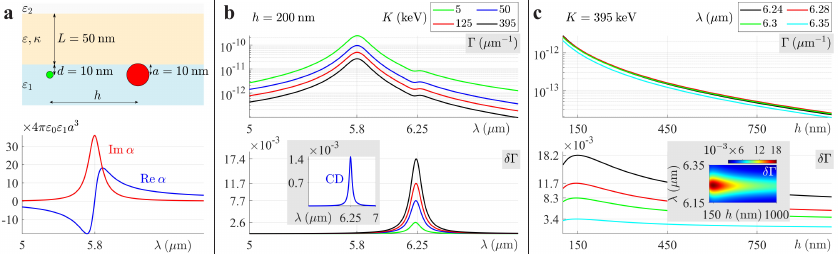}
\caption{{\bf Graphene-free chiral sensing}. {\bf a}. Setup and nanoparticle effective polarizability $\alpha$. {\bf b}. Cathodoluminescence emission probability $\Gamma$ and its dissymmetry factor $\delta \Gamma$, as functions of the wavelength, of a nanoparticle positioned at $h=200 \: {\rm nm}$ for various electron energies $K$. The slab circular dichroism $\rm CD$ is plotted in the inset for comparison purposes. {\bf c}. Dependence of $\Gamma$ and $\delta \Gamma$ on the nanoparticle position $h$, for the electron energy $K = 395 \: {\rm keV}$ at various wavelengths close to the molecular resonance. The detailed dependence of $\delta \Gamma$ on $h$ and $\lambda$ is plotted in the inset.}
\end{figure*}

\vfill \noindent {\bf CHIRAL SENSING THROUGH CATHODOLUMINESCENCE}

MOA is entirely due to molecular chirality and consequently it can be harnessed to perform chiral sensing. In our fast electron excitation scheme, this can be done by incorporating an off-axis nanoparticle located at $y=h$ in the substrate just underneath the interface, as sketched in Fig.2a, and collecting in the superstrate its cathodoluminescence emission probability $\Gamma (h)$, the number of photons emitted  per incoming electron per unit photon wavelength (see Methods and Supporting Information). Since the reflected near field is spatially asymmetric (MOA), the emission probability $\Gamma(-h)$, produced by the nanoparticle when placed in the mirror symmetric position $y=-h$, is different from $\Gamma(h)$ and hence chiral sensing efficiency is here provided by the dissymmetry factor
\begin{equation}
\delta \Gamma \left( h \right) = 2\frac{{\Gamma \left( { - h} \right) - \Gamma \left( h \right)}}{{\Gamma \left( {-  h} \right) + \Gamma \left(  h \right)}}.
\end{equation}

In Fig.3 we consider the graphene-free setup with an additional transparent conductor nanoparticle of radius $a=10 \: {\rm nm}$ whose effective polarizability $\alpha$ (nonretarded approximation) displays a plasmon resonance peak at $\lambda = 5.8 \: \mu{\rm m}$ (see Fig.3a). In Fig.3b we set the nanoparticle position $h = 200 \: {\rm nm}$ and we plot $\Gamma$ and $\delta \Gamma$ as functions of $\lambda$ for different electron energies $K$. The emission spectrum $\Gamma$ is characterized by the nanoparticle resonance peak at $5.8 \: \mu{\rm m}$ together with a slight dip due to molecular resonant absorption at $6.25 \: \mu{\rm m}$ whereas $\delta \Gamma$ only displays a peak at the molecular resonance, a strong evidence of the role played by the dissymmetry factor as enantioselectivity quantifier. Note that, higher electron energies reduce $\Gamma$ since the electron field in Eq.(\ref{Ee}) decreases for large electron velocities. Conversely, the dissymmetry factor $\delta \Gamma$ exhibits the opposite behavior since higher electron velocities increase the $\rm S$ amplitude of the electron field at the expense of its $\rm P$ amplitude (see Methods and Supporting Information) and this enhances MOA (as discussed above). For comparison purposes, in the inset of Fig.3b we have plotted the circular dichroism spectrum ${\rm CD}={\rm Im} (\kappa) k_0 L$, which turns out to be about a half of the dissymmetry factor $\delta \Gamma$ at the here considered lowest electron energy and about one order of magnitude smaller at higher energies, thus showing that the chiral sensing technique we are here discussing is more efficient than standard circular dichroism. In Fig.3c we set the electron energy $K=395 \: {\rm keV}$ and we plot $\Gamma$ and $\delta \Gamma$ as functions of the nanoparticle position $h$ at different wavelengths close to the molecular resonance. The emission spectrum $\Gamma$ fades for increasing values of $h$ due to the radial exponential decay of the electron field whereas the dissymmetry factor $\delta \Gamma$ displays a maximum at $h \simeq 150 \: {\rm nm}$ in agreement with the profile of the field dissymmetry factor $\Upsilon$ in Fig.2b. On the other hand, $\Gamma$ is poorly sensitive to $\lambda$ over the chosen spectral bandwith whereas $\delta \Gamma$ displays a marked dependence on $\lambda$ as further detailed in the inset of Fig.3c.

\begin{figure*}[!] \label{Fig4}
\includegraphics[width=1\textwidth]{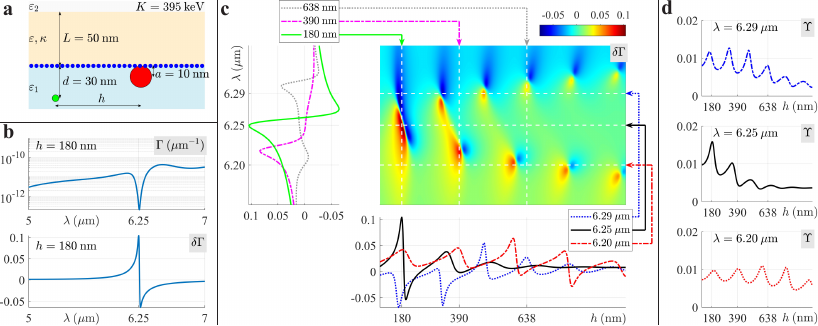}
\caption{{\bf Graphene enhanced chiral sensing}. {\bf a}. Chiral sensing setup with graphene inclusion and electron energy $K = 395 \: {\rm keV}$. {\bf b}. Cathodoluminescence emission probability $\Gamma$ and its dissymmetry factor $\delta \Gamma$, as functions of the wavelength, of a nanoparticle positioned at $h=180 \: {\rm nm}$. {\bf c} Detailed dependence of $\delta \Gamma$ on both nanoparticle position (horizontal axis) and the wavelength (vertical axis). The bottom and lateral subplots display  several slices of $\delta \Gamma$ at selected wavelengths and nanoparticle positions, respectively. {\bf d}. Field dissymmetry factors $\Upsilon(h)$ at the three wavelengths selected in the bottom plot of Fig.4c.}
\end{figure*}

The setup with both graphene inclusion and nanoparticle (see Fig.4a) is even more interesting due to the interplay of MOA and graphene plasmon polaritons excitation. We hereafter set the electron energy $K=395 \: {\rm keV}$ and in Fig.4b we report the wavelength dependence of $\Gamma$ and $\delta \Gamma$ of a nanoparticle positioned at $h = 180 \: {\rm nm}$. In this case the emission spectrum $\Gamma$ does not exhibit a peak at $\lambda = 5.8 \: \mu{\rm m}$ since nanoparticle plasmon resonance is hybridized with the broadband graphene plasmon resonance \cite{Ciat1,Ciat2}, this yielding the almost flat profile only displaying the pronounced dip at the molecular resonance $6.25 \: \mu{\rm m}$. The dissymmetry factor $\delta \Gamma$ is very sensitive to molecular chirality, with a maximum value of about $0.1$ which is two orders of magnitude larger than circular dichroism reported in the inset of Fig.3b, and its profile is not bell shaped around the molecular resonance (as in Fig.3b). The further enhancement and spectral properties of $\delta \Gamma$ are consequences of the complicated impact of graphene plasmon polaritons excitation on MOA. To discuss such effects, in Fig.4c we plot the detailed dependence of $\delta \Gamma$  on both nanoparticle position (horizontal axis) and the wavelength (vertical axis) and, in the bottom and lateral subplots, we report for clarity purposes several slices at selected wavelengths and nanoparticle positions, respectively. This plot should be compared with the inset of Fig.3 (which is its countperpart for the graphene-free setup) just to realize the dramatic impact of graphene plasmon polaritons excitation on the enantioselectvity techinque we are describing. The complicated pattern of $\delta \Gamma$ displays a modulation along $h$ whose period decreases for increasing wavelengths, an inequivocable consequence of graphene plasmon polariton dispersion $k_{\rm p}(\lambda)$ which provides marked wavelength dependence to the spatial interference pattern of the counterpropagating graphene plasmon polaritons. This is also remarked in Fig.4d where we plot the field dissymmetry factors $\Upsilon(h)$ at the three wavelengths analyzed in the bottom plot of Fig.4c, their shapes qualitatively reproducing the slices of $\delta \Gamma$.

\vfill \noindent {\bf CONCLUSIONS AND DISCUSSION} 

We have shown that the explicit symmetry breaking at the macroscopic scale provided by molecular chirality enables a geometrically symmetric chiral sample to respond in a spatially asymmetrical fashion to a mirror symmetric electromagnetic excitation. The field dissymmetry produced by the sample is particularly marked in the near field thus opening novel avenues to perform efficient chiral sensing with the aid of nanophotonic strategies. We have illustrated the concept by using a relativistic electron to trigger the asymmetric response of a chiral nanofilm which is detected through the cathodoluminescence of an off-axis nanoparticle. Chiral sensitivity turns out to be one order of magnitude larger than circular dichroism and its enhancement to two orders due to an additional graphene inclusion is a solid evidence that focused tailoring of the nanophotonic platform could in principle lead to ultra-efficient chiral sensing schemes. Therefore, the flexibility and potentials of our findings stem from the number of nowadays available nanoemitter sources (semiconductor quantum dots, fluorescent molecules, etc.) and near field optical probes (tapered optical fibers, aperture probes, solid metal tips, etc.). 

Our chiral sensing approach relies on the difference of the near field at two mirror symmetrical points produced by a single incident mirror symmetric field, whereas standard techniques based on circular dichroism exploit the different absorption of two fields of opposite circular polarization. In addition to highlight the operational difference between the two schemes, this comparison should underline the difference between their underpinning physical mechanisms since in our case the absorption rates at two symmetrical points are equal. Accordingly, in mirror optical activity, the asymmetry of the near field is sensitive to both the real and imaginary parts of the chiral parameter $\kappa$. 

Another advantage of our approach is that its setup does not require large-area microfabrication as in schemes exploiting plasmon enhanced circular dichroism \cite{Govor,Liuuu} where plasmonic inclusions are usually patterned on a surface \cite{Abdul,Neste} to increase the strength of the differential absorption signal. Remarkably our technique exploits the whole region of the homogeneous chiral nanofilm where the field is not negligible, conversely to what happens in plasmonic schemes relying on superchiral fields where the volume-averaged optical chirality is generally limited by its rapidly varying sign over the volume \cite{Schaf,Munnn}.

\vfill \noindent {\bf METHODS}

{\bf The slab field}. The field ${\bf{E}}_\omega ^{\left( \rm s \right)}$ in the chiral slab of Fig.1c satisfies Eq.(\ref{Maxwell}) whose most general solution is
\begin{equation} \label{s-field}
{\bf{E}}_\omega ^{\left(\rm s \right)}  = \int {d^2 {\bf{k}}_\parallel  } \sum\limits_{\sigma \tau } {e^{i{\bf{K}}_\tau ^{\left( \sigma  \right)}  \cdot {\bf{r}}} {\bf{V}}_\tau ^{\left( \sigma  \right)} U_\tau ^{\left( \sigma  \right)} },
\end{equation}
where
\begin{eqnarray}
{\bf{K}}_\tau ^{\left( \sigma  \right)}  &=& {\bf{k}}_\parallel   +  \sigma K_{z\tau } {\bf{e}}_z, \nonumber \\
{\bf{V}}_\tau ^{\left( \sigma  \right)}  &=& n\frac{{k_0 }}{{i\sigma K_{z\tau } }}\left( {\frac{{\kappa ^2 }}{{\sqrt {\varepsilon \kappa ^2 } }} + \tau } \right){\bf{u}}_{\rm S}  + \left( {{\bf{u}}_{\rm P}  - \frac{{k_\parallel  }}{{\sigma K_{z \tau } }}{\bf{e}}_z } \right), \nonumber \\
K_{z\tau } &=& \sqrt {k_0^2 \left( {n + \tau \kappa } \right)^2  - k_\parallel ^2}, \: \: \: \: \left({\rm Im} \: K_{z\tau } > 0\right),
\end{eqnarray}
$n = \sqrt {\varepsilon \kappa ^2 }/\kappa$, $\sigma = \pm 1$, $\tau = \pm 1$ and ${U_\tau ^{\left( \sigma  \right)} }$  are arbitrary amplitudes. The field ${\bf{E}}_\omega ^{\left(\rm s \right)}$ is the superposition of plane waves labelled by the parallel wavenumber ${\bf k}_\parallel$, with transverse wavenumber $\sigma K_{z\tau }$, and they are transverse (${\bf{K}}_\tau ^{\left( \sigma  \right)}  \cdot {\bf{V}}_\tau ^{\left( \sigma  \right)}  = 0$) and circularly polarized (${\bf{V}}_\tau ^{\left( \sigma  \right)}  \cdot {\bf{V}}_\tau ^{\left( \sigma  \right)}  = 0$) of sense $\tau$ since
\begin{eqnarray}
 {\bf{K}}_\tau ^{\left( \sigma  \right)}  \times {\bf{V}}_\tau ^{\left( \sigma  \right)}  &=& i\tau k_0 \left( {n + \tau \kappa } \right){\bf{V}}_\tau ^{\left( \sigma  \right)}.
\end{eqnarray}
As a consequence of the circular polarization state, the S and P components of ${\bf{V}}_\tau ^{\left( \sigma  \right)}$ are not independent (as opposite to plane waves in achiral media) and the ensuing effective S/P coupling is particularly strong for waves of large momentum. Indeed, for $k_\parallel \gg k_0$, the S component of ${\bf{V}}_\tau ^{\left( \sigma  \right)}$ fades due to $K_{z \tau} \simeq i k_\parallel$ in its denominator whereas the $\rm P$-$z$ part approaches the limit ${\bf{u}}_{\rm P}  + i \sigma {\bf{e}}_z$ (which is the same circular polarization state accounting for the transverse spin of evanescent waves in achiral media \cite{Fortu}). Physically, the suppression of the $\rm S$ component of large momentum evanescent waves stems from the interplay of their circular polarization state with the asymptotic circular polarization state of the $\rm P$-$z$ part (in full generality due isotropy and to transversality, $\nabla \cdot {\bf E}_\omega^{\rm (s)} = 0$, see Eq.(\ref{Maxwell})).

{\bf Reflected and transmitted fields}. The incident field ${\bf{E}}_\omega ^{\left( \rm i \right)}$ of Eq. (\ref{i-field}) in the substrate (see Fig.1c) causes the slab to produce the field ${\bf{E}}_\omega ^{\left( \rm s \right)}$ of Eq.(\ref{s-field}) in its volume and the reflected $\rm (r)$ and transmitted $\rm (t)$ fields 
\begin{eqnarray} \label{rt-fields}
{\bf{E}}_\omega ^{\left( \rm r \right)}  &=& \int {d^2 {\bf{k}}_\parallel  } e^{i{\bf{k}}_\parallel   \cdot {\bf{r}}_\parallel  }   e^{-ik_{1z} z} \left[ {U_{\rm S}^{\left( \rm r \right)} {\bf{u}}_{\rm S}  + U_{\rm P}^{\left( \rm r \right)} \left( {{\bf{u}}_{\rm P}  + \frac{{k_\parallel  }}{{k_{1z} }}{\bf{e}}_z } \right)} \right], \nonumber \\
{\bf{E}}_\omega ^{\left( \rm t \right)}  &=& \int {d^2 {\bf{k}}_\parallel  } e^{i{\bf{k}}_\parallel   \cdot {\bf{r}}_\parallel  }   e^{ik_{2z} (z-L)} \left[ {U_{\rm S}^{\left( \rm t \right)} {\bf{u}}_{\rm S}  + U_{\rm P}^{\left( \rm t \right)} \left( {{\bf{u}}_{\rm P}  - \frac{{k_\parallel  }}{{k_{2z} }}{\bf{e}}_z } \right)} \right],
\end{eqnarray}
in the substrate and superstrate, respectively, where $k_{2z}  = \sqrt {k_0^2 \varepsilon _2  - k_\parallel ^2 }$ . By enforcing the matching conditions at $z=0$ (with the current density $\sigma_{\rm G} {\bf{E}}_{\omega \parallel}$ due to graphene conductivity $\sigma_{\rm G}$, see Supporting Information) and $z=L$, a linear algebraic system for the eight unknown amplitudes of the (s),(r) and (t) fields is obtained. The reflected and transmitted field amplitudes are accordingly found to be
\begin{eqnarray}
 \left( {\begin{array}{*{20}c}
   {U_{\rm S}^{\left( \rm r \right)} }  \\
   {U_{\rm P}^{\left( \rm r \right)} }  \\
\end{array}} \right) &=& \left( {\begin{array}{*{20}c}
   {R_{\rm SS} } & {nR_{\rm SP} }  \\
   {nR_{\rm PS} } & {R_{\rm PP} }  \\
\end{array}} \right)\left( {\begin{array}{*{20}c}
   {U_{\rm S}^{\left( \rm i \right)} }  \\
   {U_{\rm P}^{\left( \rm i \right)} }  \\
\end{array}} \right), \nonumber \\ 
 \left( {\begin{array}{*{20}c}
   {U_{\rm S}^{\left( \rm t \right)} }  \\
   {U_{\rm P}^{\left( \rm t \right)} }  \\
\end{array}} \right) &=& \left( {\begin{array}{*{20}c}
   {T_{\rm SS} } & {nT_{\rm SP} }  \\
   {nT_{\rm PS} } & {T_{\rm PP} }  \\
\end{array}} \right)\left( {\begin{array}{*{20}c}
   {U_{\rm S}^{\left( \rm i \right)} }  \\
   {U_{\rm P}^{\left( \rm i \right)} }  \\
\end{array}} \right),
\end{eqnarray}
where the reflection and transmission coefficients $R_{ij}(k_\parallel)$ and $T_{ij}(k_\parallel)$ are invariant under chirality reversal $\kappa \rightarrow - \kappa$ (see Supporting Information). Since the S and P unit vectors in Fourier space satisfy the relations 
\begin{eqnarray} \label{uREF}
{\bf{u}}_{\rm S} \left( {{\mathcal{R}}{\bf{k}}_\parallel  } \right) &=&  - {\mathcal{R}}{\bf{u}}_{\rm S} \left( {{\bf{k}}_\parallel  } \right), \nonumber \\
{\bf{u}}_{\rm P} \left( {{\mathcal{R}}{\bf{k}}_\parallel  } \right) &=& {\mathcal{R}}{\bf{u}}_{\rm P} \left( {{\bf{k}}_\parallel  } \right),
\end{eqnarray}
the symmetric and antisymmetric parts of the reflected field are
\begin{eqnarray} \label{rAS}
 {\bf{E}}_\omega ^{\left( \rm r \right){\rm S}}  &=& \int {d^2 {\bf{k}}_\parallel  } e^{i{\bf{k}}_\parallel   \cdot {\bf{r}}_\parallel  }     e^{ - ik_{1z} z}  \left[ \left( {R_{\rm SS} U_{\rm S}^{\rm \left( i \right){\rm A}} + nR_{\rm SP} U_{\rm P}^{\rm \left( i \right){\rm A}} } \right){\bf{u}}_{\rm S} + \left( {nR_{\rm PS} U_{\rm S}^{\rm \left( i \right){\rm S}} + R_{\rm PP} U_{\rm P}^{\rm \left( i \right){\rm S}}} \right)\left( {{\bf{u}}_{\rm P}  + \frac{{k_\parallel  }}{{k_{1z} }}{\bf{e}}_z } \right) \right] , \nonumber \\
 {\bf{E}}_\omega ^{\left( \rm r \right){\rm A}}  &=& \int {d^2 {\bf{k}}_\parallel  } e^{i{\bf{k}}_\parallel   \cdot {\bf{r}}_\parallel  }    e^{ - ik_{1z} z} \left[ \left( {R_{\rm SS} U_{\rm S}^{\rm \left( i \right){\rm S}} + nR_{\rm SP} U_{\rm P}^{\rm \left( i \right){\rm S}}} \right){\bf{u}}_{\rm S} +  \left( {nR_{\rm PS} U_{\rm S}^{\rm \left( i \right){\rm A}} + R_{\rm PP} U_{\rm P}^{\rm \left( i \right){\rm A}}} \right)\left( {{\bf{u}}_{\rm P}  + \frac{{k_\parallel  }}{{k_{1z} }}{\bf{e}}_z } \right) \right]. \nonumber \\
\end{eqnarray}
where we have introduced the Fourier symmetric and antisymmetric parts 
\begin{eqnarray}
U_j^{\left( \rm i \right){\rm S}} \left( {{\bf{k}}_\parallel  } \right) &=& \frac{1}{2}\left[ {U_j^{\left(\rm i \right)} \left( {{\bf{k}}_\parallel  } \right) + U_j^{\left(\rm  i \right)} \left( {{\mathcal{R}}{\bf{k}}_\parallel  } \right)} \right], \nonumber \\
U_j^{\left( \rm i \right){\rm A}} \left( {{\bf{k}}_\parallel  } \right) &=& \frac{1}{2}\left[ {U_j^{\left(\rm i \right)} \left( {{\bf{k}}_\parallel  } \right) - U_j^{\left(\rm i \right)} \left( {{\mathcal{R}}{\bf{k}}_\parallel  } \right)} \right]
\end{eqnarray} 
of the incident field $j= \left\{\rm S,P \right\}$ amplitude. For an incident MSF, which is characterized by $U_{\rm S}^{\left( \rm i \right){\rm S}}=0$ and $U_{\rm P}^{\left( \rm i \right){\rm A}}=0$, Eqs.(\ref{rAS}) reduce to Eqs.(\ref{MOAr}).

{\bf Electron field}. The electron field impinges onto the chiral slab with amplitudes (see Supporting Information)
\begin{eqnarray} \label{Uee}
 U_{\rm S}^{\left( \rm ei \right)}  &=& E_{\omega 0} \frac{{e^{ik_{1z} d} }}{{2k_\parallel  }}\left( { - \frac{{k_y }}{{k_{1z} }}} \right)\delta \left( {k_x  - \frac{\omega }{v}} \right), \nonumber \\ 
 U_{\rm P}^{\left( \rm ei \right)}  &=& E_{\omega 0} \frac{{e^{ik_{1z} d} }}{{2k_\parallel  }}\left( {\frac{{k_{1z} }}{{ {\beta k_0 } \varepsilon _1 }}} \right)\delta \left( {k_x  - \frac{\omega }{v}} \right),
\end{eqnarray}
which are respectively antisymmetric and symmetric under momentum reflection $k_y \rightarrow -k_y$, as imposed by the electron field mirror symmetry. The delta function selects the photon momentum projection along the electron trajectory $k_x = \omega / v$ so that the transverse wave number is
\begin{equation} 
k_{1z}  =  i \sqrt { {\frac{\omega^2 }{{v^2\gamma^2 }}} + k_y^2 },
\end{equation}
confirming that the field spectrum is entirely evanescent in the sub-Cherenkov regime (${\rm Im} \: \gamma = 0$). Note that the ratio between the S and P amplitudes 
\begin{equation}
\frac{{\left| {U_{\rm S}^{\left(\rm ei \right)} } \right|}}{{\left| {U_{\rm P}^{\left(\rm ei \right)} } \right|}} = \frac{{\beta \varepsilon _1 }}{{\left( {\frac{1}{{\beta ^2 }} - \varepsilon _1 } \right)\frac{{k_0 }}{{\left| {k_y } \right|}} + \frac{{\left| {k_y } \right|}}{{k_0 }}}}
\end{equation}
is an increasing function of $\beta$ so that the larger the electron velocity, the more relevant the S field component.

{\bf Nanoparticle Cathodoluminescence}. We consider a transparent conductor nanoparticle of radius $a$ that we model as a point dipole (nonretarded aproximation) positioned at ${\bf{r}}_{\rm NP}  = h{\bf{e}}_y  - a{\bf{e}}_z$  whose dipole moment (in the frequency domain) is ${\bf{p}}_\omega  = \alpha {\bf{E}}_\omega ^{\left( {\rm ext} \right)}$ where ${\bf{E}}_\omega ^{\left( {\rm ext} \right)}$ is the field experienced by the dipole (without self-field) and $\alpha  = 4\pi \varepsilon _0 \varepsilon _1 a^3 \left( {\frac{{\varepsilon _{\rm NP}  - \varepsilon _1 }}{{\varepsilon _{\rm NP}  + 2\varepsilon _1 }}} \right)$ is the well-known polarizability of the sphere. The dipole field 
\begin{equation}
{\bf{E}}_\omega ^{\left( \rm p \right)}  = \left( {k_0^2 \varepsilon _1  + \nabla \nabla  \cdot } \right)\left( {\frac{1}{{4\pi \varepsilon _0 \varepsilon _1 }}\frac{{e^{ik_0 \sqrt {\varepsilon _1 } \left| {{\bf{r}} - {\bf{r}}_{\rm NP} } \right|} }}{{\left| {{\bf{r}} - {\bf{r}}_{\rm NP} } \right|}}{\bf{p}}_\omega  } \right)
\end{equation}
impinges onto the slab and produces the reflected and transmitted fields
\begin{eqnarray} \label{Eprt}
 {\bf{E}}_\omega ^{\left(\rm {pr} \right)}  &=& G^{\left(\rm {pr} \right)} {\bf{p}}_\omega  \quad \left( {z < 0} \right), \nonumber \\ 
 {\bf{E}}_\omega ^{\left(\rm {pt} \right)}  &=& G^{\left( \rm {pt} \right)} {\bf{p}}_\omega  \quad \left( {z > L} \right), 
\end{eqnarray}
where the dyadics $G^{\left(\rm {pr} \right)}$ and $G^{\left(\rm {pt} \right)}$ are expressed as Fourier integrasl containing the slab reflection and transmission coefficients, respectively (see Supporting Information). The dipole moment is self-consistently given by
\begin{equation}
{\bf{p}}_\omega   = \left\{ {\left[ {\frac{1}{\alpha }I - G^{\left( {\rm pr} \right)} } \right]^{ - 1} \left( {{\bf{E}}_\omega ^{\left(\rm e \right)}  + {\bf{E}}_\omega ^{\left( {\rm er} \right)} } \right)} \right\}_{{\bf{r}} = {\bf{r}}_{\rm NP} },
\end{equation}
where $I$ is the identity dyadic, revealing its ability to probe the electron near field. Since we are considering the sub-Cherenkov regime and the aloof electron does not produce transition radiation, cathodoluminesce radiation is only due to the nanoparticle and, in the superstrate, it can be evaluated from the far field expression 
\begin{equation} \label{farf}
{\bf{E}}_\omega ^{\left( {\rm pt} \right)}  = \frac{{e^{ik_0 \sqrt {\varepsilon _2 } r} }}{{k_0 r}}{\bf{f}}\left( {\theta ,\varphi } \right)
\end{equation}
where $\theta$ and $\varphi$ are the spherical polar angles and $\bf f$ is a vector amplitude linearly dependent on ${\bf p}_\omega$. The number of photons emitted per incoming electron, per per unit of solid angle of emission and per unit of photon wavelength is $\frac{{dN}}{{d\Omega d\lambda }} = \frac{{\lambda \sqrt \varepsilon_2  }}{{\pi \hbar Z_0 }}\left| {\bf f } \right|^2$ (see Supporting Information) so that the cathodoluminescence emission probability in the superstrate is 
\begin{equation}
\Gamma  = \frac{{\lambda \sqrt {\varepsilon _2 } }}{{\pi \hbar Z_0 }}\int\limits_{\cos \theta  > 0} {d\Omega } \left| {\bf{f}} \right|^2.
\end{equation}

%
%
%
%
%
%
%

\end{document}


\title{Mirror optical activity: efficient chiral sensing from electromagnetic parity indefiniteness. Supporting  Information}
\author{A. Ciattoni$^1$}
\email{alessandro.ciattoni@spin.cnr.it}
\affiliation{$^1$CNR-SPIN, c/o Dip.to di Scienze Fisiche e Chimiche, Via Vetoio, 67100 Coppito (L'Aquila), Italy}
\date{\today}

\begin{abstract}
This document provides supporting information to "Mirror optical activity: efficient chiral sensing from electromagnetic parity indefiniteness". We present here the discussion of the parity indefiniteness of the electromagnetic field in isotropic and homogeneous chiral media. In addition, we provide the detailed analytical description of mirror optical activity accompanying the illumination of a chiral slab by a mirror symmetric field and we specialize the general approach to the situation where the mirror symmetric field is provided by a fast electron in aloof configuration. We conclude with the analytical description of the mirror optical activity detection thorough the cathodoluminescence radiation produced by a nanoparticle.  
\end{abstract}

\maketitle
\renewcommand\theequation{S\arabic{equation}}
\renewcommand\thefigure{S\arabic{figure}}
\renewcommand\thesection{S\arabic{section}}
\renewcommand\thesubsection{S\arabic{section}.\arabic{subsection}}
\count\footins = 1000

\section{Homogeneous Chiral Media}
The constitutive relations modelling (in the frequency domain)  the electromagnetic response of a homogenous and isotropic non-magnetic chiral medium are
\begin{eqnarray} \label{chi_const}
 {\bf{D}}_\omega   &=& \varepsilon _0 \varepsilon {\bf{E}}_\omega   - \frac{i}{c}\kappa {\bf{H}}_\omega, \nonumber   \\ 
 {\bf{B}}_\omega   &=& \mu _0 {\bf{H}}_\omega   + \frac{i}{c}\kappa {\bf{E}}_\omega,
\end{eqnarray}
where $\varepsilon _0$ and $\mu _0$ are the vacuum permittivity and permeability, respectively, $c = \frac{1}{\sqrt{\varepsilon _0 \mu _0}}$ is the speed of light, $\varepsilon$ is the relative permittivity  and $\kappa$ is the Pasteur parameter. The subscript $\omega$ hereafter labels a quantity in the frequency domain according to the spectral analysis
\begin{equation}
f_\omega   = \frac{1}{{2\pi }}\int_{ - \infty }^{ + \infty } {dt} \;e^{i\omega t} f\left( t \right).
\end{equation}
The parameter $\kappa$ controls the coupling between electric and magnetic responses of the medium and, therefore, the strength of its chirality. In the relevant case of a molecular medium where the chiral molecules are dispersed in a dielectric matrix medium, the permittivity and chirality parameter can be modelled as
\begin{eqnarray} \label{chi_disp}
 \varepsilon  &=& \varepsilon _{\rm b}  - \gamma_{\rm mol} \left( {\frac{1}{{\hbar \omega  - \hbar \omega _0  + i\Gamma }} - \frac{1}{{\hbar \omega  + \hbar \omega _0  + i\Gamma }}} \right), \nonumber \\ 
 \kappa  &=& \beta_{\rm mol} \left( {\frac{1}{{\hbar \omega  - \hbar \omega _0  + i\Gamma }} + \frac{1}{{\hbar \omega  + \hbar \omega _0  + i\Gamma }}} \right) 
\end{eqnarray}
where $\varepsilon _b$ is the background refractive index and the coefficients $\gamma_{\rm mol}$ and $\beta_{\rm mol}$ are the amplitudes of absorptive and chiral properties; $\omega_0 = 2\pi c / \lambda_0$ corresponds to the wavelength $\lambda_0$ of a molecular absorption resonance with broadening determined by the damping $\Gamma$. Relations (\ref{chi_disp}) can be derived from the quantum equation of motion for the density matrix of a chiral molecule \cite{Govor1} whose lack of definite parity enables the coupling between its electric and magnetic dipole moments in turn responsible for the magneto-electric coupling contained in Eqs.(\ref{chi_const}).

In the present work we focus on chiroptical phenomena occurring in the infrared and we have chosen alanine as a specific molecular species for numerical calculations since it exhibits marked vibrational circular dichroism \cite{Tulioo,Jahnig}. Accordingly we hereafter set $\varepsilon_{\rm b} = 1.3$, $\rm \gamma_{mol} = 9.8 \cdot 10^{-4} \: eV$, $\rm \beta_{mol} = 4.68 \cdot 10^{-5} \: eV$, $\rm \hbar \omega_0 = 1.98 \cdot 10^{-1} \: eV$ (corresponding to the absorption resonance wavelength $\rm \lambda_0 = 6.25 \: \mu m$) and $\rm \Gamma = 1.6 \cdot 10^{-3} \: eV$ in order to emulate the resonant chiral response of an alanine enantiomer resulting from its vibrational mode at $\rm 1600 \: cm^{-1}$ (the two enantiomers have $\beta_{\rm mol}$, and hence $\kappa$, with opposite signs).

\section{Parity indefiniteness of the electromagnetic field in chiral media}
An object is chiral if it is distinguishable from its mirror image or, in other words, if its mirror image cannot be rigidly superposed onto it by only resorting to rotations and translation. Such geometric definition is based on the mirror reflection transformation which consequently plays a pivotal role in the analysis of any physical effect produced by chirality. Since in this work we focus on homogeneous isotropic chiral media, without loss of generality we will consider the (active) reflection trough the $xz$ plane given by 
\begin{equation}
{\bf{r}}' = \left( {{\bf{e}}_x {\bf{e}}_x - {\bf{e}}_y {\bf{e}}_y  + {\bf{e}}_z {\bf{e}}_z } \right){\bf{r}} \equiv {\mathcal{R}} {\bf{r}}
\end{equation}
where ${\bf{e}}_x$, ${\bf{e}}_y$ and ${\bf{e}}_z$ are the standard cartesian unit vectors and the dyadic notation $( {{\bf{ab}}} ){\bf{c}} = \left( {{\bf{b}} \cdot {\bf{c}}} \right) {\bf{a}}$ has be used. Basically the dyadic ${\mathcal{R}}$ change the sign of the $y-$component of the vector it operates on. While the mirror image of a geometric object $O$ is the object $O'$ whose points ${\bf{r}}'$ are obtained from the points ${\bf{r}}$ of $O$ through ${\bf{r}}' = {\mathcal{R}} {\bf{r}}$, each physical entity has its own behavior under reflections. The most relevant behaviors are those of polar vectors $\bf P ({\bf r})$, axial vectors $\bf A ({\bf r})$, scalars $s ({\bf r})$ and pseudoscalars $p ({\bf r})$ whose mirror images are ${\bf{P}}'\left( {\bf{r}} \right) = {\mathcal{R}}{\bf{P}}\left( {{\mathcal{R}}{\bf{r}}} \right)$, ${\bf{A}}'\left( {\bf{r}} \right) =  - {\mathcal{R}_y}{\bf{A}}\left( {{\mathcal{R}}{\bf{r}}} \right)$, $s'\left( {\bf{r}} \right) = s\left( {{\mathcal{R}}{\bf{r}}} \right)$ and $p'\left( {\bf{r}} \right) =  - p\left( {{\mathcal{R}}{\bf{r}}} \right)$.

Reflection invariance (or mirror symmetry) is among the fundamental symmetries of nature, and it states that if a complete experiment is subjected to mirror reflection, the resulting experiment should, in principle, be realizable (when the weak interaction can be neglected). Maxwell's equations $\nabla  \times {\bf{E}}_\omega   = i\omega {\bf{B}}_\omega$ and $\nabla  \times {\bf{H}}_\omega   =  - i\omega {\bf{D}}_\omega$ evidently have mirror symmetry since they are left invariant by the transformation ${\bf{r}}' = {\mathcal{R}} {\bf{r}}$ combined with the requirement that ${\bf{E}}_\omega, {\bf{D}}_\omega$ are polar vectors and ${\bf{H}}_\omega, {\bf{B}}_\omega$ are axial vectors 
\footnote{Mathematically, this is a consequence of the relation $\nabla  \times \left[ {{\mathcal{R}}{\bf{F}}\left( {{\mathcal{R}}{\bf{r}}} \right)} \right] =  - {\mathcal{R}}\left[ {\left( {\nabla  \times {\bf{F}}} \right)\left( {{\mathcal{R}}{\bf{r}}} \right)} \right]$ holding for any vector field $\bf{F}({\bf{r}})$.}
. Enforcing these properties on the constitutive relations of Eqs.(\ref{chi_const}) it is straightforward proving that, under reflection, $\varepsilon ' = \varepsilon$ and $\kappa' = - \kappa$, i.e. the permittivity $\varepsilon$ is a scalar and the chiral parameter $\kappa$ is a pseudoscalar. Since opposite molecular enantiomers provide opposite chiral parameters, it follows that the mirror image of a chiral medium is precisely its opposite enantiomeric medium, as expected. As a consequence, reflection invariance can here be rephrased as: the mirror image of the electromagnetic field existing in a chiral medium is a physically realizable electromagnetic field inside the opposite enantiomeric medium. This well-known fact is the starting point of the following analysis and hence it is convenient to mathematically summarize it by stating that if the fields ${\bf{E}}_\omega, {\bf{H}}_\omega$ satisfy Maxwell's equations with chiral parameter $\kappa$, the mirror images ${\bf{E}'}_\omega, {\bf{H}'}_\omega$ satisfy Maxwell's equations with chiral parameter $-\kappa$, and viceversa, i.e.
\begin{equation} \label{RI}
\left\{ \begin{array}{@{\mkern0mu} l}
\displaystyle \nabla  \times {\bf{E}}_\omega   = i\omega \left( {\mu _0 {\bf{H}}_\omega   + \frac{i}{c}\kappa {\bf{E}}_\omega  } \right) \\ 
\displaystyle \nabla  \times {\bf{H}}_\omega   = i\omega \left( { - \varepsilon _0 \varepsilon {\bf{E}}_\omega   + \frac{i}{c}\kappa {\bf{H}}_\omega  } \right) \\ 
 \end{array} \right. 
\quad \quad \mathop  \Leftrightarrow \limits_{\def\arraystretch{1.2} \begin{array}{c}
    \uparrow   \\
   {\bf{E}}'_\omega  \left( {\bf{r}} \right) = {\mathcal{R}}{\bf{E}}_\omega  \left( {{\mathcal{R}}{\bf{r}}} \right) \\
   {\bf{H}}'_\omega  \left( {\bf{r}} \right) =  - {\mathcal{R}}{\bf{H}}_\omega  \left( {{\mathcal{R}}{\bf{r}}} \right)  \\
\end{array}} 
\quad \quad \left\{ \begin{array}{@{\mkern0mu} l}
\displaystyle \nabla  \times {\bf{E}}'_\omega   = i\omega \left( {\mu _0 {\bf{H}}'_\omega   - \frac{i}{c}\kappa {\bf{E}}'_\omega  } \right) \\ 
\displaystyle \nabla  \times {\bf{H}}'_\omega   = i\omega \left( { - \varepsilon _0 \varepsilon {\bf{E}}'_\omega   - \frac{i}{c}\kappa {\bf{H}}'_\omega  } \right) \\ 
 \end{array} \right. .
\end{equation}

As any fundamental symmetry, reflection invariance of physical laws only holds for a ''complete experiment'', i.e. for an isolated system. In our case the isolated system is composed of the electromagnetic field and the chiral medium whose global mirror reflection leaves Maxwell's equations invariant. What is remarkable here is that reflection invariance also provides a spatial dissymmetry property of the single electromagnetic field existing in a chiral medium. In order to discuss this point it is convenient to consider the concept of mirror symmetric field (MSF) that is an electromagnetic field such that ${\bf{E}}_\omega, {\bf{H}}_\omega$ are identical to their mirror images ${\bf{E}}_\omega = {\bf{E}'}_\omega$, ${\bf{H}}_\omega = {\bf{H}'}_\omega$ or equivalenty satisfying the relations
\begin{eqnarray} \label{MSF}
 {\bf{E}}_\omega  \left( {\bf{r}} \right) &=& {\mathcal{R}} {\bf{E}}_\omega  \left( {{\mathcal{R}} {\bf{r}}} \right), \nonumber \\ 
 {\bf{H}}_\omega  \left( {\bf{r}} \right) &=&  - {\mathcal{R}} {\bf{H}}_\omega  \left( {{\mathcal{R}} {\bf{r}}} \right). 
 \end{eqnarray}
Now an electromagnetic field in a chiral medium is such that ${\bf{E}}_\omega, {\bf{H}}_\omega$ satisfy the left system of Eqs.(\ref{RI}) and hence, by reflection invariance, their mirror images ${\bf{E}}'_\omega, {\bf{H}}'_\omega$ satisfy the right system which is different from the left one due to the sign reversal of $\kappa$. Consequently the relations ${\bf{E}}_\omega = {\bf{E}'}_\omega$, ${\bf{H}}_\omega = {\bf{H}'}_\omega$ are forbidden by chirality and this proves that MSFs do not exist in chiral media. 
In other words, in a homogeneous and isotropic chiral medium, the microscopic lack of mirror symmetry of chiral molecules is responsible for a macroscopic spatial dissymmetry of the electromagnetic field. The peculiarity of such spatial dissymmetry becomes even more evident by noting that, conversely, MSFs are allowed in achiral media. As a matter of fact, by repeating the above reasoning in the case $\kappa = 0$, it is evident that the mirror images ${\bf{E}}'_\omega$, ${\bf{H}}'_\omega$ satisfy the same equations as ${\bf{E}}_\omega$, ${\bf{H}}_\omega$ so that MSFs in achiral media are not ruled out. As an example, the electromagnetic field produced by an electric dipole ${\bf{p}}_\omega$ in a achiral medium is easily seen to be a MSF (with respect to the reflection $\mathcal{R}$) if the dipole moment lies on the $xz$ plane (i.e. if ${\bf{e}}_y  \cdot {\bf{p}}_\omega   = 0$).

Mirror dissymmetry of the electromagnetic field in chiral media can also be analyzed in a different way which sheds further light upon such fundamental trait. After solving for the magnetic fields ${\bf{H}}_\omega   = \frac{1}{{i\omega \mu _0 }}\left( {\nabla  \times {\bf{E}}_\omega   + k_0 \kappa {\bf{E}}_\omega  } \right)$ and ${\bf{H}}'_\omega   = \frac{1}{{i\omega \mu _0 }}\left( {\nabla  \times {\bf{E}}'_\omega   - k_0 \kappa {\bf{E}}'_\omega  } \right)$ from the curl equations for ${\bf{E}}_\omega$ and ${\bf{E}'}_\omega$ in Eqs.(\ref{RI}), the curl equations for ${\bf{H}}_\omega$ and ${\bf{H}'}_\omega$ yield
\begin{eqnarray} \label{Maxw_E_E'}
 \left[ {k_0^2 \left( {\varepsilon  - \kappa ^2 } \right) - \nabla  \times \nabla  \times } \right]{\bf{E}}_\omega   - 2k_0 \kappa \nabla  \times {\bf{E}}_\omega   &=& 0, \nonumber  \\ 
 \left[ {k_0^2 \left( {\varepsilon  - \kappa ^2 } \right) - \nabla  \times \nabla  \times } \right]{\bf{E}}'_\omega   + 2k_0 \kappa \nabla  \times {\bf{E}}'_\omega   &=& 0, 
\end{eqnarray}
where $k_0 = \omega/c$ is the vacuum wavenumber. Consider now the electric field decomposition ${\bf{E}}_\omega   = {\bf{E}}_\omega ^{\rm S }  + {\bf{E}}_\omega ^{\rm A }$, where the fields
\begin{eqnarray} \label{SymAsym}
 {\bf{E}}_\omega ^{\rm S }  &=& \frac{1}{2}\left( {{\bf{E}}_\omega   + {\bf{E}}'_\omega  } \right), \nonumber  \\ 
 {\bf{E}}_\omega ^{\rm A }  &=& \frac{1}{2}\left( {{\bf{E}}_\omega   - {\bf{E}}'_\omega  } \right), 
\end{eqnarray}
have definite parity since ${\bf{E}}_\omega ^{\rm S }$ is even (symmetric part) and ${\bf{E}}_\omega ^{\rm A }$ is odd (antisymmetric part) under reflection. By definition a MSF is such that ${\bf{E}}_\omega   = {\bf{E}}_\omega ^{\rm S }$ (i.e. ${\bf{E}}_\omega ^{\rm A }  = 0$) whereas a mirror antisymmetric field (MAF) is characterized by the condition ${\bf{E}}_\omega   = {\bf{E}}_\omega ^{\rm A }$ (i.e. ${\bf{E}}_\omega ^{\rm S }  = 0$). From Eqs.(\ref{Maxw_E_E'}) it is straightforward to obtain the equations
\begin{eqnarray} \label{SA}
 \left[ {k_0^2 \left( {\varepsilon  - \kappa ^2 } \right) - \nabla  \times \nabla  \times } \right]{\bf{E}}_\omega ^{\rm S }  &=& 2k_0 \kappa \nabla  \times {\bf{E}}_\omega ^{\rm A }, \nonumber  \\ 
 \left[ {k_0^2 \left( {\varepsilon  - \kappa ^2 } \right) - \nabla  \times \nabla  \times } \right]{\bf{E}}_\omega ^{\rm A }  &=& 2k_0 \kappa \nabla  \times {\bf{E}}_\omega ^{\rm S },
\end{eqnarray}
showing that the medium chirality $\kappa$ physically couples the symmetric and the antisymmetric parts of the electric field. The structure of such coupling clearly reveals that ${\bf{E}}_\omega ^{\rm S }$ and ${\bf{E}}_\omega ^{\rm A }$ can not separately vanish or, in other words that the chiral medium hosts neither MSFs (as just proven above) nor MAFs. We conclude that the field has always its symmetric and antisymmetric parts and therefore it is an indefinite parity field (IPF). Note that in achiral media, Eqs.(\ref{SA}) with $\kappa = 0$ show that the symmetric and the antisymmetric parts are not coupled and hence both MSFs and MAFs are separately allowed so that electromagnetic parity indefiniteness does not occur.

\section{Mirror Optical Activity}
As discussed in the above section, the electromagnetic field in a chiral medium  necessarily has indefinite parity (IPF) whereas in a achiral medium the field can have definite parity (MSFs or MAFs). It is therefore interesting to investigate the effect of a MSF impinging onto a chiral slab  since an IPF is expected to show up in the surrounding achiral environment as a consequence of the IPF excited inside the  bulk of the chiral slab. Most remarkably, the dissymmetry of the reflected (and transmitted) field is a genuinely novel chiro-optical effect, the mirror optical activity (MOA), since it is entirely due to the slab chirality and its detection effectively amounts to a measurement of the chiral parameter $\kappa$.

\begin{figure*} \label{FigS1}
\includegraphics[width=0.7\textwidth]{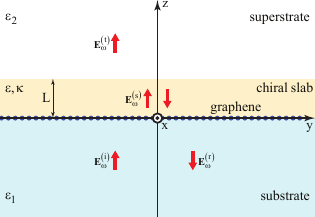}
\caption{Geometry of the chiral slab surrounded by an achiral environment made up of a substrate, a graphene sheet and a dielectric. Red arrows sketch the field configuration.}
\end{figure*}

\subsection{Interaction of a chiral slab with an electromagnetic field}
As a prelude to the discussion of MOA it is essential to revise the full response of a chiral slab to an impinging electromagnetic field. With reference to Fig.S1, we consider a chiral homogeneous and isotropic medium of permittivity $\varepsilon$ and chiral parameter $\kappa$ (see Sec.S1) filling the slab $0<z<L$ surrounded by a substrate in $z<0$ and a superstrate in $z>L$ of permittivities $\varepsilon_1$ and $\varepsilon_2$, respectively. In order to consider a slightly more general situation, we also incorporare a graphene sheet lying on the $z=0$ plane between the substrate and the chiral slab. Even though graphene is not essential for the onset of MOA, it is nonetheless remarkable that the excitation of plasmon polaritons combined with field dissymmetry triggers a novel near field interferometric mechanism able to enhance the enantiomeric sensing (see below). 

Due to the planar geometry of the setup, we resort to the angular spectrum representation technique and the incident field ${\bf{E}}_\omega ^{\left( \rm i \right)}$ in $z<0$ is
\begin{equation} \label{i-field}
{\bf{E}}_\omega ^{\left( \rm i \right)}  = \int {d^2 {\bf{k}}_\parallel  } e^{i{\bf{k}}_\parallel   \cdot {\bf{r}}_\parallel  } e^{ik_{1z} z} \left[ {U_{\rm S}^{\left( \rm i \right)} {\bf{u}}_{\rm S}  + U_{\rm P}^{\left( \rm i \right)} \left( {{\bf{u}}_{\rm P}  - \frac{{k_\parallel  }}{{k_{1z} }}{\bf{e}}_z } \right)} \right]
\end{equation}
where ${\bf{r}}_\parallel   = x {\bf{e}}_x  + y {\bf{e}}_y$, ${\bf{k}}_\parallel   = k_x {\bf{e}}_x  + k_y {\bf{e}}_y$ is the parallel wavenumber, $k_{1z}  = \sqrt {k_0^2 \varepsilon _1  - k_\parallel ^2 }$ (${\rm Im} \: k_{1z} > 0$), ${\bf{u}}_{\rm S} \left( {{\bf{k}}_\parallel  } \right) = {\bf{e}}_z  \times \frac{{{\bf{k}}_\parallel  }}{{k_\parallel  }}$,  $ {\bf{u}}_{\rm P} \left( {{\bf{k}}_\parallel  } \right) = \frac{{{\bf{k}}_\parallel  }}{{k_\parallel  }}$ are the transverse electric (S) and transverse magnetic (P) unit vectors and $U_{\rm S}^{\left( \rm i \right)}  ( {{\bf{k}}_\parallel})$, $U_{\rm P}^{\left( \rm i \right)} ( {{\bf{k}}_\parallel})$ are the S and P spectral amplitudes of the incident $(\rm i)$ field
\footnote{The unit vectors ${\bf{u}}_{\rm S}$, ${\bf{u}}_{\rm P}$ are an orthonormal basis of the parallel plane of vectors ${\bf{A}}_\parallel   = A_x {\bf{e}}_x  + A_y {\bf{e}}_y$. The ordered triple  $\left({\bf{u}}_{\rm P},{\bf{u}}_{\rm S},{\bf e}_z \right)$ is a left-handed orthonormal basis since ${\bf{u}}_{\rm P}  \times {\bf{u}}_{\rm S} = {\bf{e}}_z$, ${\bf{e}}_z  \times {\bf{u}}_{\rm P} = {\bf{u}}_{\rm S}$, $ {\bf{u}}_{\rm S}  \times {\bf{e}}_z = {\bf{u}}_{\rm P}$. The $j=\{\rm S,P\}$ component of the incident field is given by $U_j^{\left( \rm i \right)} ({\bf k}_\parallel) = {\bf{u}}_j ({\bf k}_\parallel) \cdot \int {\frac{{d^2 {\bf{r}}_\parallel  }}{{\left( {2\pi } \right)^2 }}}  e^{ - i{\bf{k}}_\parallel   \cdot {\bf{r}}_\parallel  } {\bf{E}}_\omega ^{\left( \rm i \right)} \left( {{\bf{r}}_\parallel   + 0^ -  {\bf{e}}_z } \right)$. 
}
. The reflected field ${\bf{E}}_\omega ^{\left( \rm r \right)}$ in $z<0$ and the transmitted field ${\bf{E}}_\omega ^{\left( \rm t \right)}$ in $z>L$ are
\begin{eqnarray} \label{rt-fields}
{\bf{E}}_\omega ^{\left( \rm r \right)}  &=& \int {d^2 {\bf{k}}_\parallel  } e^{i{\bf{k}}_\parallel   \cdot {\bf{r}}_\parallel  } e^{-ik_{1z} z} \left[ {U_{\rm S}^{\left( \rm r \right)} {\bf{u}}_{\rm S}  + U_{\rm P}^{\left( \rm r \right)} \left( {{\bf{u}}_{\rm P}  + \frac{{k_\parallel  }}{{k_{1z} }}{\bf{e}}_z } \right)} \right], \nonumber \\
{\bf{E}}_\omega ^{\left( \rm t \right)}  &=& \int {d^2 {\bf{k}}_\parallel  } e^{i{\bf{k}}_\parallel   \cdot {\bf{r}}_\parallel  } e^{ik_{2z} (z-L)} \left[ {U_{\rm S}^{\left( \rm t \right)} {\bf{u}}_{\rm S}  + U_{\rm P}^{\left( \rm t \right)} \left( {{\bf{u}}_{\rm P}  - \frac{{k_\parallel  }}{{k_{2z} }}{\bf{e}}_z } \right)} \right],
\end{eqnarray}
where $k_{2z}  = \sqrt {k_0^2 \varepsilon _2  - k_\parallel ^2 }$ (${\rm Im} \: k_{2z} > 0$) and $U_{\rm S}^{\left( \rm r \right)}  ( {{\bf{k}}_\parallel})$, $U_{\rm P}^{\left( \rm r \right)} ( {{\bf{k}}_\parallel})$, $U_{\rm S}^{\left( \rm t \right)}  ( {{\bf{k}}_\parallel})$, $U_{\rm P}^{\left( \rm t \right)} ( {{\bf{k}}_\parallel})$ are the S and P spectral amplitudes of the reflected $(\rm r)$ and  transmitted $(\rm t)$ fields. The field ${\bf{E}}_\omega ^{\left( \rm s \right)}$ in $0<z<L$ is the most general solution of the first of Eqs.(\ref{Maxw_E_E'}) given by
\begin{equation} \label{s-field}
{\bf{E}}_\omega ^{\left( \rm s \right)}  = \int {d^2 {\bf{k}}_\parallel  } e^{i{\bf{k}}_\parallel   \cdot {\bf{r}}_\parallel  } \sum\limits_{\sigma \tau } {e^{i\sigma K_{z\tau } z} \left[ { n {\frac{{ k_0 }}{{i \sigma K_{z\tau } }}} \left( {\frac{{\kappa ^2 }}{{\sqrt {\varepsilon \kappa ^2 } }} + \tau } \right) {\bf{u}}_{\rm S}  +\left( {\bf{u}}_{\rm P}  - \frac{{k_\parallel  }}{{\sigma K_{z\tau } }}{\bf{e}}_z \right) }  \right]U_\tau ^{\left( \sigma  \right)} }, 
\end{equation}
where $n = \frac{\sqrt {\varepsilon \kappa ^2 }}{\kappa}$ and it is the superposition of circularly polarized plane waves whose spectral amplitudes are  ${U_\tau ^{\left( \sigma  \right)} } ( {{\bf{k}}_\parallel})$ and they are labeled by the parallel wavenumber ${\bf{k}}_\parallel$, the sign $\sigma = \pm 1$ of the transverse wavenumber $K_{z\tau } = \sqrt {k_0^2 \left( {n + \tau \kappa } \right)^2  - k_\parallel ^2 }$ (${\rm Im} \: K_{z\tau } > 0$) and the circular polarization sense $\tau= \pm 1$. Indeed, the wavevector ${\bf{K}}_\tau ^{\left( \sigma  \right)}  = {\bf{k}}_\parallel   +  \sigma K_{z\tau } {\bf{e}}_z$ and the polarization ${\bf{V}}_\tau ^{\left( \sigma  \right)}  = n\frac{{k_0 }}{{i\sigma K_{z\tau } }}\left( {\frac{{\kappa ^2 }}{{\sqrt {\varepsilon \kappa ^2 } }} + \tau } \right){\bf{u}}_{\rm S}  + \left( {{\bf{u}}_{\rm P}  - \frac{{k_\parallel  }}{{\sigma K_{z\tau } }}{\bf{e}}_z } \right)$ of the basic plane waves are easily seen to satisfy the relation
\begin{equation}
 {\bf{K}}_\tau ^{\left( \sigma  \right)}  \times {\bf{V}}_\tau ^{\left( \sigma  \right)}  = i\tau k_0 \left( {n + \tau \kappa } \right){\bf{V}}_\tau ^{\left( \sigma  \right)}  
\end{equation}
showing their transversality (${\bf{K}}_\tau ^{\left( \sigma  \right)}  \cdot {\bf{V}}_\tau ^{\left( \sigma  \right)}=0$) and circular polarization (${\bf{V}}_\tau ^{\left( \sigma  \right)}  \cdot {\bf{V}}_\tau ^{\left( \sigma  \right)}  = 0$) with sense $\tau$. The state of circular polarization imposes a severe constraint  between the S and P components of the field in Eq.(\ref{s-field}). Such S/P coupling is entirely produced by chirality and it is particularly strong along the evanescent part of the spectrum. Indeed, for $k_\parallel \gg k_0$, the S component of ${\bf{V}}_\tau ^{\left( \sigma  \right)}$ fades due to $K_{z \tau} \simeq i k_\parallel$ in its denominator whereas the P component approaches its limit ${\bf{u}}_{\rm P}  + i\sigma {\bf{e}}_z$ (which is the same circular polarization state accounting for the transverse spin of evanescent waves in achiral media \cite{Fortuno}). The fact that the evanescent part of the spectrum is almost entirely P is physically clear since isotropy of the chiral medium together with  field transversality ($\nabla \cdot {\bf E}_\omega^{\rm (s)} = 0$, see the first of Eqs.(\ref{Maxw_E_E'})) imply that the $\rm P$-$\rm z$ part is asymptotically circularly polarized and this forces the asymptotic vanishing behavior of the orthogonal S component to preserve the  circular  polarization state of the overall plane wave.
  
Graphene is modeled through a surface current density flowing along the plane $z=0$ and given by ${\bf{K}}_{\omega  \parallel } = \sigma_{\rm G} (\omega) {\bf{E}}_{\omega  \parallel }$ where $\sigma_{\rm G} (\omega)$ is the graphene surface conductivity (see Sec.S6) and ${\bf{E}}_{\omega  \parallel } = E_{\omega x} {\bf e}_x + E_{\omega y} {\bf e}_y$ is the in-plane part of total electric field at $z=0$. By enforcing the boundary conditions at $z=0$ and $z=L$ 
\begin{equation} 
\def\arraystretch{1.6}
\left\{ \begin{array}{@{\mkern0mu} l @{\mkern0mu}}
\textstyle  {\bf{E}}_{\omega \parallel } \left( {{\bf{r}}_\parallel   + 0^ -  {\bf{e}}_z } \right) - {\bf{E}}_{\omega \parallel } \left( {{\bf{r}}_\parallel   + 0^ +  {\bf{e}}_z } \right) = 0 \\ 
 \displaystyle  \left[ {{\bf{H}}_\omega  \left( {{\bf{r}}_\parallel   + 0^ -  {\bf{e}}_z } \right) - {\bf{H}}_\omega  \left( {{\bf{r}}_\parallel   + 0^ +  {\bf{e}}_z } \right)} \right] \times {\bf{e}}_z  = \sigma _{\rm G} {\bf{E}}_{\omega \parallel } \left( {{\bf{r}}_\parallel   + 0{\bf{e}}_z } \right) \\ 
 \end{array} \right.,  \: \:
\left\{ \begin{array}{@{\mkern0mu} l @{\mkern0mu}}
 {\bf{E}}_{\omega \parallel } \left( {{\bf{r}}_\parallel   + L^ -  {\bf{e}}_z } \right) - {\bf{E}}_{\omega \parallel } \left( {{\bf{r}}_\parallel   + L^ +  {\bf{e}}_z } \right) = {\bf{0}} \\ 
 \left[ {{\bf{H}}_\omega  \left( {{\bf{r}}_\parallel   + L^ -  {\bf{e}}_z } \right) - {\bf{H}}_\omega  \left( {{\bf{r}}_\parallel   + L^ +  {\bf{e}}_z } \right)} \right] \times {\bf{e}}_z  = 0 \\ 
 \end{array} \right. ,
\end{equation}
we obtain 
\begin{equation} \label{equa}
\def\arraystretch{2.5}
 \left\{ \begin{array}{@{\mkern0mu} l @{\mkern0mu}}
 \displaystyle \sum\limits_{\sigma \tau } {\left( {\frac{{k_0 }}{{2ik_{1z} }}} \right)\left[ {\varepsilon  + \frac{{\varepsilon  + \tau \sqrt {\varepsilon \kappa ^2 } }}{{\sigma K_{z\tau } }}\left( {k_{1z}  + k_0 Z_0 \sigma _{\rm G} } \right)} \right]\tau U_\tau ^{\left( \sigma  \right)} }  = nU_{\rm S}^{\left( \rm i \right)}  \\ 
\displaystyle \sum\limits_{\sigma \tau } {e^{i\sigma K_{z\tau } L} \left( {1 - \frac{{\varepsilon  + \tau \sqrt {\varepsilon \kappa ^2 } }}{{\sigma K_{z\tau } }}k_{2z} } \right)\tau U_\tau ^{\left( \sigma  \right)} }  = 0 \\ 
 \displaystyle \sum\limits_{\sigma \tau } {\left( {\frac{{k_{1z} }}{{2\varepsilon _1 }}} \right)\left( {\frac{{\varepsilon  + \tau \sqrt {\varepsilon \kappa ^2 } }}{{\sigma K_{z\tau } }} + \frac{{\varepsilon _1 }}{{k_{1z} }} + \frac{{Z_0 \sigma _{\rm G} }}{{k_0 }}} \right)U_\tau ^{\left( \sigma  \right)} }  = U_{\rm P}^{\left( \rm i \right)}  \\ 
\displaystyle \sum\limits_{\sigma \tau } {e^{i\sigma K_{z\tau } L} \left( {\varepsilon _2  - \frac{{\varepsilon  + \tau \sqrt {\varepsilon \kappa ^2 } }}{{\sigma K_{z\tau } }}k_{2z} } \right)U_\tau ^{\left( \sigma  \right)} }  = 0 \\ 
 \end{array} \right., \quad
\left\{ \begin{array}{@{\mkern0mu} l @{\mkern0mu}}
\displaystyle U_{\rm S}^{\left( \rm r \right)}  =  - U_{\rm S}^{\left( \rm i \right)}  + n\sum\limits_{\sigma \tau } {\left[ {\frac{{k_0 }}{{i\sigma K_{z\tau } }}\left( {\frac{{\kappa ^2 }}{{\sqrt {\varepsilon \kappa ^2 } }} + \tau } \right)} \right]U_\tau ^{\left( \sigma  \right)} }  \\ 
\displaystyle U_{\rm P}^{\left( \rm r \right)}  =  - U_{\rm P}^{\left( \rm i \right)}  + \sum\limits_{\sigma \tau } {U_\tau ^{\left( \sigma  \right)} }  \\ 
\displaystyle U_{\rm S}^{\left( \rm t \right)}  = n\sum\limits_{\sigma \tau } {e^{i\sigma K_{z\tau } L} \left[ {\frac{{k_0 }}{{i\sigma K_{z\tau } }}\left( {\frac{{\kappa ^2 }}{{\sqrt {\varepsilon \kappa ^2 } }} + \tau } \right)} \right]U_\tau ^{\left( \sigma  \right)} }  \\ 
\displaystyle U_{\rm P}^{\left( \rm t \right)}  = \sum\limits_{\sigma \tau } {e^{i\sigma K_{z\tau } L} U_\tau ^{\left( \sigma  \right)} }  \\ 
 \end{array} \right. ,
\end{equation}
where $Z_0 = \sqrt{\frac{\mu_0}{ \varepsilon_0}}$ is the vacuum impedance. In matrix notation, Eqs.(\ref{equa}) can be casted as
\begin{equation} \label{equa_matr}
\def\arraystretch{2.0}
\underline{\underline M} \: \underline U  = nU_{\rm S}^{\left( \rm i \right)} \underline V _{\rm S}  + U_{\rm P}^{\left( \rm i \right)} \underline V _{\rm P} ,\quad \quad \left\{ \begin{array}{@{\mkern0mu} l @{\mkern0mu}}
U_{\rm S}^{\left( \rm r \right)}  =  - U_{\rm S}^{\left( \rm i \right)}  + n {\underline W _{\rm S}^{\left( \rm r \right)} }  \cdot \underline U  \\ 
 U_{\rm P}^{\left( \rm r \right)}  =  - U_{\rm P}^{\left( \rm i \right)}  +  {\underline W _{\rm P}^{\left( \rm r \right)} } \cdot \underline U  \\ 
 U_{\rm S}^{\left( \rm t \right)}  = n {\underline W _{\rm S}^{\left( \rm t \right)} } \cdot \underline U  \\ 
 U_{\rm P}^{\left( \rm t \right)}  =  {\underline W _{\rm P}^{\left( \rm t \right)} } \cdot \underline U  \\ 
 \end{array} \right. ,
\end{equation}
where $\underline{\underline M}$ is the system matrix,
$\underline U  = \begin{pmatrix} {U_ + ^{\left(  +  \right)} } & {U_ - ^{\left(  +  \right)} } & {U_ + ^{\left(  -  \right)} } & {U_ - ^{\left(  -  \right)} }  \end{pmatrix}^{\rm T}$,  $\underline V _{\rm S}  = \begin{pmatrix} 1 & 0 & 0 & 0  \end{pmatrix}^{\rm T}$, $\underline V _{\rm P}  = \begin{pmatrix} 0 & 0 & 1 & 0  \end{pmatrix}^{\rm T}$ and the vectors $\underline W _{\rm S}^{\left(\rm r \right)} ,\underline W _{\rm P}^{\left( \rm r \right)} ,\underline W _{\rm S}^{\left(\rm t \right)} ,\underline W _{\rm P}^{\left( \rm t \right)}$ are easily read from the right system of Eqs.(\ref{equa}).  Substituting the formal solution $\underline U  = \underline{\underline M} ^{ - 1} \left( {nU_{\rm S}^{\left(\rm i \right)} \underline V _{\rm S}  + U_{\rm P}^{\left(\rm i \right)} \underline V _{\rm P} } \right)$ of the left system of Eqs.(\ref{equa_matr}) into  the right system we obtain
\begin{equation} \label{solu}
\def\arraystretch{2.0}
\left\{ \begin{array}{@{\mkern0mu} l @{\mkern0mu}}
 \begin{pmatrix}
   {U_{\rm S}^{\left(\rm r \right)} }  \\
   {U_{\rm P}^{\left(\rm r \right)} }  \\
\end{pmatrix}= \begin{pmatrix}
   {R_{\rm SS} } & {nR_{\rm SP} }  \\
   {nR_{\rm PS} } & {R_{\rm PP} }  \\
\end{pmatrix} \begin{pmatrix}
   {U_{\rm S}^{\left(\rm i \right)} }  \\
   {U_{\rm P}^{\left(\rm i \right)} }  \\
\end{pmatrix}\\ 
 \begin{pmatrix}
   {U_{\rm S}^{\left(\rm t \right)} }  \\
   {U_{\rm P}^{\left(\rm t \right)} }  \\
\end{pmatrix}= \begin{pmatrix}
   {T_{\rm SS} } & {nT_{\rm SP} }  \\
   {nT_{\rm PS} } & {T_{\rm PP} }  \\
\end{pmatrix} \begin{pmatrix}
   {U_{\rm S}^{\left(\rm i \right)} }  \\
   {U_{\rm P}^{\left(\rm i \right)} }  \\
\end{pmatrix}\\ 
 \end{array} \right., \: \: {\rm where} \: \:
\left\{ \begin{array}{@{\mkern0mu} l @{\mkern0mu}}
 \begin{pmatrix}
   {R_{\rm SS} } & {R_{\rm SP} }  \\
   {R_{\rm PS} } & {R_{\rm PP} }  \\
\end{pmatrix}= \begin{pmatrix}
   { - 1 + n^2 \underline W _{\rm S}^{\left(\rm r \right)}  \cdot \underline{\underline M} ^{ - 1} \underline V _{\rm S} } & {\underline W _{\rm S}^{\left(\rm r \right)}  \cdot \underline{\underline M} ^{ - 1} \underline V _{\rm P} }  \\
   {\underline W _{\rm P}^{\left(\rm r \right)}  \cdot \underline{\underline M} ^{ - 1} \underline V _{\rm S} } & { - 1 + \underline W _{\rm P}^{\left(\rm r \right)}  \cdot \underline{\underline M} ^{ - 1} \underline V _{\rm P} }  \\
\end{pmatrix}\\ 
 \begin{pmatrix}
   {T_{\rm SS} } & {T_{\rm SP} }  \\
   {T_{\rm PS} } & {T_{\rm PP} }  \\
\end{pmatrix}= \begin{pmatrix}
   {n^2 \underline W _{\rm S}^{\left(\rm t \right)}  \cdot \underline{\underline M} ^{ - 1} \underline V _{\rm S} } & {\underline W _{\rm S}^{\left(\rm t \right)}  \cdot \underline{\underline M} ^{ - 1} \underline V _{\rm P} }  \\
   {\underline W _{\rm P}^{\left(\rm t \right)}  \cdot \underline{\underline M} ^{ - 1} \underline V _{\rm S} } & {\underline W _{\rm P}^{\left(\rm t \right)}  \cdot \underline{\underline M} ^{ - 1} \underline V _{\rm P} }  \\
\end{pmatrix}  
 \end{array} \right.
\end{equation}
which directly provide the reflected and transmitted amplitudes in terms of the incident ones. 
\begin{figure*} \label{FigS2}
\includegraphics[width=1\textwidth]{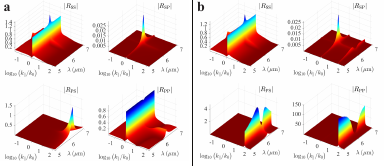}
\caption{Reflection and Transmission coefficients of a chiral slab as functions of the normalized parallel wavevector $k_\parallel / k_0$ and of the wavelength $\lambda = 2 \pi c/\omega$, without ($\bf a$) and with ($\bf b$) graphene sheet.}
\end{figure*}
Inserting these amplitudes into Eqs.(\ref{rt-fields}) we eventually obtain 
\begin{eqnarray} \label{react}
 {\bf{E}}_\omega ^{\left(\rm r \right)}  &=& \int {d^2 {\bf{k}}_\parallel  } e^{i{\bf{k}}_\parallel   \cdot {\bf{r}}_\parallel  } e^{ - ik_{1z} z} \left[ {\left( {R_{\rm SS} U_{\rm S}^{\left(\rm i \right)}  + nR_{\rm SP} U_{\rm P}^{\left(\rm i \right)} } \right){\bf{u}}_{\rm S}  + \left( {nR_{\rm PS} U_{\rm S}^{\left(\rm i \right)}  + R_{\rm PP} U_{\rm P}^{\left(\rm i \right)} } \right)\left( {{\bf{u}}_{\rm P}  + \frac{{k_\parallel  }}{{k_{1z} }}{\bf{ e}}_z } \right)} \right], \nonumber \\ 
 {\bf{E}}_\omega ^{\left(\rm t \right)}  &=& \int {d^2 {\bf{k}}_\parallel  } e^{i{\bf{k}}_\parallel   \cdot {\bf{r}}_\parallel  } e^{ik_{2z} \left( {z - L} \right)} \left[ {\left( {T_{\rm SS} U_{\rm S}^{\left(\rm i \right)}  + nT_{\rm SP} U_{\rm P}^{\left(\rm i \right)} } \right){\bf{u}}_{\rm S}  + \left( {nT_{\rm PS} U_{\rm S}^{\left(\rm i \right)}  + T_{\rm PP} U_{\rm P}^{\left(\rm i \right)} } \right)\left( {{\bf{u}}_{\rm P}  - \frac{{k_\parallel  }}{{k_{2z} }}{\bf{ e}}_z } \right)} \right].
\end{eqnarray}
which yield the reflected and transmitted fields in terms of the amplitudes of the incident field and hence they provide the full-electromagnetic response of the chiral slab. Note that each of the S and P components of the reflected and transmitted field depends on both $U_{\rm S}^{\left(\rm i \right)}$, $U_{\rm P}^{\left(\rm i \right)}$ as a consequence of the S/P coupling produced by slab chirality. Since in achiral media the S and P component are uncoupled, the mixing coefficients $R_{\rm SP}$, $R_{\rm PS}$, $T_{\rm SP}$, $T_{\rm PS}$ are entirely due to chirality and they are essential in the theoretical discussion of MOA (see below). It is worth noting that reflection $R$ and transmission $T$ matrices in Eqs.(\ref{solu}) are left invariant by the reflection ${\bf k}_\parallel \rightarrow {\mathcal{R}} {\bf k}_\parallel$ and by chirality reversal $\kappa \rightarrow -\kappa$ since $n^2 = \varepsilon$ and, from Eqs.(\ref{equa}), the matrix $\underline{\underline M}$ and the vectors $\underline{W}$ only depend on  $k_\parallel = \sqrt{k_x^2 + k_y^2}$ and $\kappa^2$. 

In Fig.S2a we plot the reflection coefficients, as functions of the normalized parallel wavevector $k_\parallel / k_0$ and the wavelength $\lambda = 2 \pi c/\omega$, characterizing a chiral slab of thickness $L= 0.05 \: {\rm \mu m}$, filled by the molecular medium discussed in Sec.S1 and surrounded by a substrate of permittivity $\varepsilon_1 = 1.2$ and by vacuum as superstrate $\varepsilon_2 = 1$ (without graphene sheet). The molecular resonance peak at $\lambda = 6.25 \: {\rm \mu m}$ is evident together with its peculiar impact on $R_{\rm PS}$ and $R_{\rm PP}$ along the evanescent part of the spectrum. This is due to the dominant P character of the spectrum of the slab field  for $k_\parallel \gg k_0$ (as discussed above) and, accordingly, the reflected field spectrum is asymptotically almost P. Remarkably the fundamental mixing coefficient $R_{\rm PS}$ increases with $k_\parallel$. This is a consequence of the fact that the mixing reflection process where a photon of kind $S$ is reflected in a photon a kind $P$ is extremely sensitive to the incident magnetic field 
\begin{equation} \label{i-Hfield}
{\bf{H}}_\omega ^{\left(\rm i \right)}  = \frac{1}{{i\omega \mu _0 }}\nabla  \times {\bf{E}}_\omega ^{\left(\rm i \right)} = \frac{1}{{Z_0 }}\int {d^2 {\bf{k}}_\parallel  } e^{i{\bf{k}}_\parallel   \cdot {\bf{r}}_\parallel  } e^{ik_{1z} z} \left[ {\frac{{k_0 }}{{k_{1z} }}\varepsilon _1 U_{\rm P}^{\left(\rm  i \right)} {\bf{u}}_{\rm S}  - \frac{{k_{1z} }}{{k_0 }}U_{\rm S}^{\left(\rm  i \right)} \left( {{\bf{u}}_{\rm P}  - \frac{{k_\parallel  }}{{k_{1z} }}{\bf{e}}_z } \right)} \right]
\end{equation}
and the longitudinal wavenumber $k_{1z}$ multiplying $U_{\rm S}^{\left(\rm  i \right)}$ is responsible for the boosting of $R_{\rm PS}$ for $k_\parallel \gg k_0$. In Fig.S2b we plot the coefficients of the same setup but encompassing a graphene sheet of Fermi energy $E_{\rm F} = 0.5 \: {\rm eV}$ and the occurrence of the graphene plasmon resonance is manifest along the stripe  $30 \: k_0 < k_\parallel < 40 \: k_0$ (except for the purely S coefficient $R_{\rm SS}$, as expected).

\subsection{Field parity analysis}
We start our theoretical parity analysis from the slab field ${\bf{E}}_\omega ^{\left(\rm s \right)}$ of Eq.(\ref{s-field}). After noting that the S and P unit vectors in Fourier space satisfy the relations 
\begin{eqnarray} \label{uREF}
{\bf{u}}_{\rm S} \left( {{\mathcal{R}}{\bf{k}}_\parallel  } \right) &=&  - {\mathcal{R}}{\bf{u}}_{\rm S} \left( {{\bf{k}}_\parallel  } \right), \nonumber \\
{\bf{u}}_{\rm P} \left( {{\mathcal{R}}{\bf{k}}_\parallel  } \right) &=& {\mathcal{R}}{\bf{u}}_{\rm P} \left( {{\bf{k}}_\parallel  } \right),
\end{eqnarray}
and using the relation $K_{z\tau } ( {\mathcal{R}} {\bf k}_\parallel ) = K_{z\tau } ( {\bf k}_\parallel )$, it is straightforward proving that the mirror image of the slab field ${{\bf{E}}_\omega ^{\left( \rm s \right)}}$ is 
\begin{equation} \label{s-mirror}
{ {{\bf{E}}_\omega ^{\left( \rm s \right)} }}'\left( {\bf{r}} \right) = \int {d^2 {\bf{k}}_\parallel  } e^{i{\bf{k}}_\parallel   \cdot {\bf{r}}_\parallel  } \sum\limits_{\sigma \tau } {e^{i\sigma K_{z\tau } z} \left[ { - n\frac{{k_0 }}{{i\sigma K_{z\tau } }}\left( {\frac{{\kappa ^2 }}{{\sqrt {\varepsilon \kappa ^2 } }} + \tau } \right){\bf{u}}_{\rm S}  + \left( {{\bf{u}}_{\rm P}  - \frac{{k_\parallel  }}{{\sigma K_{z\tau } }}{\bf{e}}_z } \right)} \right]{ U_\tau ^{\left( \sigma  \right)} \left( {{\mathcal{R}}{\bf{k}}_\parallel  } \right) }} .
\end{equation}
The mirror field in Eq.(\ref{s-mirror}) has the same structure of the slab field of Eq.(\ref{s-field}), the only difference being the minus sign in front of $n$ so that, since all the other contributions only depends on $\kappa ^2$, it is evident that the field ${ {{\bf{E}}_\omega ^{\left( \rm s \right)} } }'$ satisfies the second of Eqs.(\ref{Maxw_E_E'}). In addition to explicitly provide a check of the full electromagnetic reflection invariance (as discussed in Sec.S2), this reasoning clarifies that, due to the S/P coupling in a  chiral medium, the fields in two opposite enantiomers only differ in the relative sign of the S and P components and that taking the mirror image is equivalent to switch the sign of the S component.

We now evaluate the symmmetric and antisymmetric parts, defined in Eqs.(\ref{SymAsym}), of the fields produced by the slab in the achiral environment using the angular spectrum representations obtained in the above subsection. The mirror image of the incident field ${\bf{E}}_\omega ^{\left( \rm i \right)}$ (see Eq.(\ref{i-field})) is
\begin{equation}
{{{\bf{E}}_\omega ^{\left( \rm i \right)} } }'  = \int {d^2 {\bf{k}}_\parallel  } e^{i{\bf{k}}_\parallel   \cdot {\bf{r}}_\parallel  } e^{ik_{1z} z} \left[ { - U_{\rm S}^{\left( \rm i \right)} \left( {{\mathcal{R}} {\bf{k}}_\parallel  } \right){\bf{u}}_{\rm S}  + U_{\rm P}^{\left( \rm i \right)} \left( {{\mathcal{R}}{\bf{k}}_\parallel  } \right)\left( {{\bf{u}}_{\rm P}  - \frac{{k_\parallel  }}{{k_{1z} }}{\bf{e}}_z } \right)} \right],
\end{equation}
where we have used Eqs.(\ref{uREF}), and consequently
\begin{eqnarray}
 {\bf{E}}_\omega ^{\left( \rm i \right){\rm S}}  &=& \int {d^2 {\bf{k}}_\parallel  } e^{i{\bf{k}}_\parallel   \cdot {\bf{r}}_\parallel  } e^{ik_{1z} z} \left[ {U_{\rm S}^{\left( \rm i \right){\rm A}} {\bf{u}}_{\rm S}  + U_{\rm P}^{\left( \rm i \right){\rm S}} \left( {{\bf{u}}_{\rm P}  - \frac{{k_\parallel  }}{{k_{1z} }}{\bf{e}}_z } \right)} \right], \nonumber \\ 
 {\bf{E}}_\omega ^{\left( \rm i \right){\rm A}}  &=& \int {d^2 {\bf{k}}_\parallel  } e^{i{\bf{k}}_\parallel   \cdot {\bf{r}}_\parallel  } e^{ik_{1z} z} \left[ {U_{\rm S}^{\left( \rm i \right){\rm S}} {\bf{u}}_{\rm S}  + U_{\rm P}^{\left( \rm i \right){\rm A}} \left( {{\bf{u}}_{\rm P}  - \frac{{k_\parallel  }}{{k_{1z} }}{\bf{e}}_z } \right)} \right], 
\end{eqnarray}
where we have introduced the symmetric and antisymmetric parts 
\begin{eqnarray}
U_j^{\left( \rm i \right){\rm S}} \left( {{\bf{k}}_\parallel  } \right) &=& \frac{1}{2}\left[ {U_j^{\left(\rm i \right)} \left( {{\bf{k}}_\parallel  } \right) + U_j^{\left(\rm  i \right)} \left( {{\mathcal{R}}{\bf{k}}_\parallel  } \right)} \right], \nonumber \\
U_j^{\left( \rm i \right){\rm A}} \left( {{\bf{k}}_\parallel  } \right) &=& \frac{1}{2}\left[ {U_j^{\left(\rm i \right)} \left( {{\bf{k}}_\parallel  } \right) - U_j^{\left(\rm i \right)} \left( {{\mathcal{R}}{\bf{k}}_\parallel  } \right)} \right]
\end{eqnarray} 
of the $j= \left\{\rm S,P \right\}$ amplitudes. Such decomposition directly provides a momentum space characterization of incident MSFs, since the condition ${\bf{E}}_\omega ^{\left( \rm i \right){\rm A}} = 0$ (see Sec.S2) is equivalent to $U_{\rm S}^{\left( \rm i \right){\rm S}}=U_{\rm P}^{\left( \rm i \right){\rm A}}=0$. In plain English, the incident field is a MSF if its S and P amplitudes are antisymmetric and symmetric, respectively, i.e. 
\begin{eqnarray} \label{UMSF}
U_{\rm S}^{\left( \rm i \right)} ( {{\bf{k}}_\parallel  } ) &=&  - U_{\rm S}^{\left( \rm i \right)} ( {{\mathcal{R}}{\bf{k}}_\parallel  } ), \nonumber \\
U_{\rm P}^{\left( \rm i \right)} ( {{\bf{k}}_\parallel  } ) &=&   U_{\rm P}^{\left( \rm i \right)} ( {{\mathcal{R}}{\bf{k}}_\parallel  } ).
\end{eqnarray}
Analogously an incident fields is a MAF if it is has symmetric S and antisymmetric P spectral amplitudes. Since there is no constraint on the field spectral amplitudes in a achiral medium, the incident field can have definite parity (MSF and MAF) or not (IPF), in agreement with the discussion at the end of Sec.S2. The mirror image of the reflected field ${\bf{E}}_\omega ^{\left( \rm r \right)}$ (see the first of Eqs.(\ref{react})) is
\begin{eqnarray}
{ {{\bf{E}}_\omega ^{\left( \rm r \right)} } }' = \int {d^2 {\bf{k}}_\parallel  } e^{i{\bf{k}}_\parallel   \cdot {\bf{r}}_\parallel  }  e^{ - ik_{1z} z} && \left\{  - \left[ {R_{\rm SS} U_{\rm S}^{\left(\rm i \right)} \left( {{\mathcal{R}} {\bf{k}}_\parallel  } \right) + nR_{\rm SP} U_{\rm P}^{\left(\rm i \right)} \left( {{\mathcal{R}} {\bf{k}}_\parallel  } \right)} \right]{\bf{u}}_{\rm S}  +  \right. \nonumber \\ 
&& \left.  + \left[ {nR_{\rm PS} U_{\rm S}^{\left(\rm i \right)} \left( {{\mathcal{R}} {\bf{k}}_\parallel  } \right) + R_{\rm PP} U_{\rm P}^{\left(\rm i \right)} \left( {{\mathcal{R}} {\bf{k}}_\parallel  } \right)} \right]\left( {{\bf{u}}_{\rm P}  + \frac{{k_\parallel  }}{{k_{1z} }}{\bf{e}}_z } \right) \right\}
\end{eqnarray}
where we have used Eqs.(\ref{uREF}) and the invariance of the reflection matrix $R$ under the change ${\bf k}_\parallel \rightarrow {\mathcal{R}} {\bf k}_\parallel$, so that consequently 

\begin{eqnarray} \label{rAS}
 {\bf{E}}_\omega ^{\left( \rm r \right){\rm S}}  &=& \int {d^2 {\bf{k}}_\parallel  } e^{i{\bf{k}}_\parallel   \cdot {\bf{r}}_\parallel  } e^{ - ik_{1z} z} \left[ {\left( {R_{\rm SS} U_{\rm S}^{\rm \left( i \right){\rm A}} + nR_{\rm SP} U_{\rm P}^{\rm \left( i \right){\rm A}} } \right){\bf{u}}_{\rm S}  + \left( {nR_{\rm PS} U_{\rm S}^{\rm \left( i \right){\rm S}} + R_{\rm PP} U_{\rm P}^{\rm \left( i \right){\rm S}}} \right)\left( {{\bf{u}}_{\rm P}  + \frac{{k_\parallel  }}{{k_{1z} }}{\bf{e}}_z } \right)} \right], \nonumber \\
 {\bf{E}}_\omega ^{\left( \rm r \right){\rm A}}  &=& \int {d^2 {\bf{k}}_\parallel  } e^{i{\bf{k}}_\parallel   \cdot {\bf{r}}_\parallel  } e^{ - ik_{1z} z} \left[ {\left( {R_{\rm SS} U_{\rm S}^{\rm \left( i \right){\rm S}} + nR_{\rm SP} U_{\rm P}^{\rm \left( i \right){\rm S}}} \right){\bf{u}}_{\rm S}  + \left( {nR_{\rm PS} U_{\rm S}^{\rm \left( i \right){\rm A}} + R_{\rm PP} U_{\rm P}^{\rm \left( i \right){\rm A}}} \right)\left( {{\bf{u}}_{\rm P}  + \frac{{k_\parallel  }}{{k_{1z} }}{\bf{e}}_z } \right)} \right]. \nonumber \\
\end{eqnarray}
The most striking feature of Eqs.(\ref{rAS}) is that $ {\bf{E}}_\omega ^{\left( \rm r \right){\rm S}}$ and ${\bf{E}}_\omega ^{\left( \rm r \right){\rm A}}$ separately vanish only for very specific incident fields whose spectral amplitudes depend on the reflection coefficients $R_{ij}$ so that we conclude that the reflected field is generally a IPF. Analogous considerations holds for the transmitted field ${\bf{E}}_\omega ^{\left( \rm t \right)}$.

\begin{figure*} \label{FigS3}
\includegraphics[width=1\textwidth]{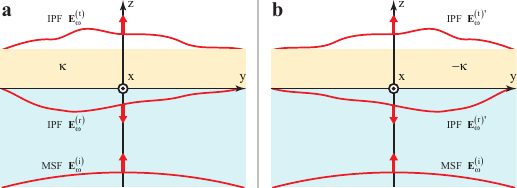}
\caption{{\bf a} Mirror optical activity (MOA): generation of reflected and transmitted IPFs out of chiral slab in the presence of an incident MSF. {\bf b} The reflected and transmitted field generated by the opposite enantiomeric slab (with chirality parameter $-\kappa$) are the mirror images of the corresponding fields in panel {\bf a}.}
\end{figure*}

\subsection{MOA}
The situation where the incident field is a MSF is particularly relevant since in this case, exploiting Eqs.(\ref{UMSF}), Eqs.(\ref{rAS}) turn into
\begin{eqnarray} \label{MOAr}
 {\bf{E}}_\omega ^{\left( \rm r \right){\rm S}}  &=& \int {d^2 {\bf{k}}_\parallel  } e^{i{\bf{k}}_\parallel   \cdot {\bf{r}}_\parallel  } e^{ - ik_{1z} z} \left[ {R_{\rm SS} U_{\rm S}^{\left( \rm i \right)} {\bf{u}}_{\rm S}  + R_{\rm PP} U_{\rm P}^{\left( \rm i \right)} \left( {{\bf{u}}_{\rm P}  + \frac{{k_\parallel  }}{{k_{1z} }}{\bf{e}}_z } \right)} \right], \nonumber  \\ 
 {\bf{E}}_\omega ^{\left( \rm r \right){\rm A}}  &=& n\int {d^2 {\bf{k}}_\parallel  } e^{i{\bf{k}}_\parallel   \cdot {\bf{r}}_\parallel  } e^{ - ik_{1z} z} \left[ {R_{\rm SP} U_{\rm P}^{\left( \rm i \right)} {\bf{u}}_{\rm S}  + R_{\rm PS} U_{\rm S}^{\left( \rm i \right)} \left( {{\bf{u}}_{\rm P}  + \frac{{k_\parallel  }}{{k_{1z} }}{\bf{e}}_z } \right)} \right] ,
\end{eqnarray}
revealing that the reflected field necessarily is an IPF since both its symmetric and its antisymmetric parts vanish if and only if the incident field vanish. The antisymmetric part ${\bf{E}}_\omega ^{\left( \rm r \right){\rm A}}$ contains only the mixing reflection coefficients $R_{\rm SP}$ and $R_{\rm PS}$ which are entirely produced by the slab chirality (see the above subsection). We conclude that the generation of reflected  and transmitted  IPFs (${\bf{E}}_\omega ^{\left( \rm r \right)}$ and  ${\bf{E}}_\omega ^{\left( \rm t \right)}$) out of chiral slab in the presence of an incident MSF (${\bf{E}}_\omega ^{\left( \rm i \right)}$) is a genuine chiro-optical effect which we refer to as mirror optical activity (MOA). As an additional remark, note that the S and P amplitudes of ${\bf{E}}_\omega ^{\left( \rm r \right){\rm A}}$ in the second of Eqs.(\ref{MOAr}) are proportional to the P and S amplitudes of ${\bf{E}}_\omega ^{\left( \rm i \right)}$, respectively. Such S/P dependency switch pertaining MOA fully agrees with the very basic electric-magnetic coupling (see Eqs.(\ref{chi_const})) lying at the heart of every chiro-optical effect since the incident magnetic field of Eq.(\ref{i-Hfield}) exhibits the same S/P dependency switch as the antisymmetric part of the electric field.

Since the basic feature of MOA is the dissymmetry of the reflected (and transmitted) field, a local estimation of the process efficiency is provided by the ratio $\Upsilon = {\left| {{\bf{E}}_\omega ^{\rm A} } \right|}/{\left| {{\bf{E}}_\omega ^{\rm S} } \right|}$ between the moduli of the antisymmetric and symmetric parts of the overall field  ${\bf{E}}_\omega   = {\bf{E}}_\omega ^{\left(\rm i \right)}  + {\bf{E}}_\omega ^{\left(\rm r \right)}$. In the substrate ($z<0$) it reduces to 
\begin{equation} \label{dissF}
\Upsilon = \frac{{\left| {{\bf{E}}_\omega ^{\left(\rm r \right){\rm A}} } \right|}}{{\left| {{\bf{E}}_\omega ^{\left(\rm i \right)}  + {\bf{E}}_\omega ^{\left(\rm  r \right){\rm S}} } \right|}}.
\end{equation}
so that, in order to observe efficient MOA, the antisimmetric part ${{\bf{E}}_\omega ^{\rm (r) A} }$ should be enhanced at the expense of the symmetric part ${{\bf{E}}_\omega ^{\rm (r) S} }$. Since the mixing coefficient $R_{\rm PS}$ is prominent along the evanescent spectrum where $R_{\rm SP}$ and $R_{\rm SS}$ are usually negligible (see Fig.S2), Eqs.(\ref{MOAr}) reveal that MOA is expected to be marked when the incident field has a prevalent S character with marked subwavelength features (so that, at the same time, ${R_{\rm PS} U_{\rm S}^{\left(\rm i \right)} }$ is enhanced and ${R_{\rm PP} U_{\rm P}^{\left(\rm  i \right)} }$ is reduced while  ${R_{\rm SS} U_{\rm S}^{\left(\rm i \right)} }$ remains small).

In Fig.S3a we sketch the MOA phenomenology in order to highlight the dissymmetry of both reflected and transmitted fields. A particularly interesting feature of MOA is that if the same incident MSF ${\bf{E}}_\omega ^{\left( \rm i \right)}$ is launched onto a second identical slab but filled with the opposite enantiomeric medium (with chirality parameter $-\kappa$), the reflected and transmitted fields are ${ {{\bf{E}}_\omega ^{\left( \rm r \right)} } }'$ and ${ {{\bf{E}}_\omega ^{\left( \rm t \right)} } }' $, i.e. they are the mirror images of the fields ${ {{\bf{E}}_\omega ^{\left( \rm r \right)} } }$ and ${ {{\bf{E}}_\omega ^{\left( \rm t \right)} } }$ produced by the first slab, as sketched in Fig.S3b. Indeed, using the invariance of the $R$ coefficients under chirality reversal $\kappa \rightarrow -\kappa$, we obtain from Eqs.(\ref{MOAr}) that the symmetric and antisymmetric parts of the field reflected by the second slab are ${\bf{E}}_\omega ^{\left( \rm r \right){\rm S}}$ and $-{\bf{E}}_\omega ^{\left( \rm r \right){\rm A}}$, the latter minus sign coming from sign flip of $n$ in the second of Eqs.(\ref{MOAr}), and these are precisely the symmetric and antisymmetric parts of the mirror image field ${ {{\bf{E}}_\omega ^{\left( \rm r \right)} } }'$. Such MOA feature suggests viable enantiometric sensing techniques where the detection of the reflected and transmitted fields allows the discrimination of different enantiomerically pure samples or the measurement of the enantiomeric excess of a mixture.

\section{MOA triggered by fast electrons}
As discussed in the above section, the basic requirement for observing MOA is the excitation of the chiral slab by means of a MSF. In addition, since the S/P coupling supporting MOA is particularly marked along the evanescent spectrum, it is high desirable to choose an incident MSF with distinct subwavelength spatial features. The field generated by a fast electron meets both such requirements and hence it is among the best available choices for the incident field. In addition the electron velocity remarkably affects the mutual S/P field content in momentum space and hence it can be exploited to efficiently tune MOA phenomenology.

An electron of charge $-e<0$, moving at constant velocity $v>0$ along the trajectory ${\bf{r}}^{\left( \rm e \right)} \left( t \right) = vt {\bf{e}}_x  - d {\bf{e}}_z$ in a transparent and unbounded dielectric of permittivity $\varepsilon_1$, generates the field \cite{deAbajo} (for $\omega >0$)
\begin{eqnarray} \label{Ee}
 {\bf{E}}_\omega ^{\left( \rm e \right)}  &=& \frac{{E_{\omega 0} }}{{\varepsilon _1 \beta ^2 \gamma }}e^{i\frac{\omega }{v}x} \left[ { - K_1 \left( {\frac{{\omega R}}{{v\gamma }}} \right){\bf{e}}_R  + \frac{i}{\gamma }K_0 \left( {\frac{{\omega R}}{{v\gamma }}} \right){\bf{e}}_x } \right], \nonumber \\ 
 {\bf{H}}_\omega ^{\left( \rm e \right)}  &=& \frac{{E_{\omega 0} }}{{Z_0 \beta \gamma }}e^{i\frac{\omega }{v}x} \left[ { - K_1 \left( {\frac{{\omega R}}{{v\gamma }}} \right){\bf{e}}_\Phi  } \right],
\end{eqnarray}
where $E_{\omega 0}  = \frac{{eZ_0 \omega }}{{4\pi ^2 c}}$ is a field amplitude, $\beta = \frac{v}{c}$, $\gamma  = \frac{1}{\sqrt{1 - \varepsilon_1 \beta ^2}}$ is the Lorentz contraction factor and $K_n$ are the modified Bessel function of the second kind. Here cylindrical coordinates $(R,\Phi)$ coaxial with the charge trajectory have been introduced according to $y = R\cos \Phi$, $z =  - d + R\sin \Phi$ and ${\bf{e}}_R  = \nabla R$ and ${\bf{e}}_\Phi   = {\bf{e}}_z  \times {\bf{e}}_R$ are the radial and azimuthal unit vectors. The field in Eqs.(\ref{Ee}) is rotationally invariant around the electron trajectory with the electric field lying on the radial $Rx$ plane and the magnetic field being purely azimuthal. Such field is a MSF since it satisfies Eqs.(\ref{MSF}) as it is straightforwardly checked by resorting to the relations $R\left( {{\mathcal{R}}{\bf{r}}} \right) = R\left({\bf{r}} \right)$, ${\bf{e}}_R \left( {{\mathcal{R}}{\bf{r}}} \right) = {\mathcal{R}}{\bf{e}}_R \left( {\bf{r}} \right)$, ${\bf{e}}_\Phi \left( {{\mathcal{R}}{\bf{r}}} \right) = - {\mathcal{R}}{\bf{e}}_\Phi \left( {\bf{r}} \right)$ and ${\bf{e}}_z  = {\mathcal{R}}{\bf{e}}_z$. In addition, the field displays marked subwavelength features in the sub-Cherenkov regime $v< \frac{c}{\sqrt \varepsilon_1}$ (where ${\mathop{\rm Im}\nolimits} \: \gamma  = 0$) since it exponentially decays along the radial direction (through the modified Bessel functions) and it does not provide electromagnetic radiation. 

The evaluation of the electron field in momentum space is simplified by resorting to the Hertz vector ${\bf \Pi} _\omega ^{\left( \rm e \right)}$ formalism 

\begin{equation}  \label{Pe}
{\bf \Pi} _\omega ^{\left( \rm e \right)}  = \frac{{E_{\omega 0} }}{{ik_0^2 \varepsilon _1 }}e^{i\frac{\omega }{v}x} K_0 \left( {\frac{{\omega R}}{{v\gamma }}}\right){\bf{e}}_x ,\quad \quad \left\{ \begin{array}{l}
 {\bf{E}}_\omega ^{\left( \rm e \right)}  = \left( {k_0^2 \varepsilon _1  + \nabla \nabla  \cdot } \right){\bf \Pi} _\omega ^{\left( \rm e \right)}  \\ 
 {\bf{H}}_\omega ^{\left( \rm e \right)}  = \frac{{k_0 \varepsilon _1 }}{{iZ_0 }}\nabla  \times {\bf \Pi} _\omega ^{\left( \rm e \right)}  \\ 
 \end{array} \right.,
\end{equation}
since ${\bf \Pi} _\omega ^{\left( \rm e \right)}$ is everywhere along the $x-$ axis. By using the relation
\begin{equation}
K_0 \left( {h\sqrt {a^2  + b^2 } } \right) = \frac{1}{2}\int_{ - \infty }^{ + \infty } {dk} \: e^{ika} \frac{{e^{ - \sqrt {h^2  + k^2 } \left| b \right|} }}{{\sqrt {h^2  + k^2 } }}
\end{equation}
for the modified Bessel function \cite{Schwinger} and noting that $\frac{{\omega ^2 }}{{v^2 \gamma ^2 }} =  - k_0^2 \varepsilon _1 + \frac{{\omega ^2 }}{{v^2 }}$, we obtain from the first of Eqs.(\ref{Pe})
\begin{equation}
{\bf \Pi} ^{\left( \rm e \right)}_\omega  = \int {d^2 {\bf{k}}_\parallel  } e^{i{\bf{k}}_\parallel   \cdot {\bf{r}}_\parallel  } e^{ik_{1z} \left| {z + d} \right|}  {\frac{{E_{\omega 0} }}{{2k_0^2 \varepsilon _1 k_{1z} }} \delta \left( {k_x  - \frac{\omega }{v}} \right){\bf{e}}_x } 
\end{equation}
where the delta function has been introduced to select the longitudinal wavenumber $\omega/v$. Evaluating the electric field from such Fourier representation of the Hertz vector by using the first equation in the right system of Eqs.(\ref{Pe}) we get
\begin{equation} \label{FourierE}
{\bf{E}}_\omega ^{\left( \rm e \right)}  = \int {d^2 {\bf{k}}_\parallel  } e^{i{\bf{k}}_\parallel   \cdot {\bf{r}}_\parallel  } e^{ik_{1z} \left| {z + d} \right|} \frac{{E_{\omega 0} }}{{2k_\parallel  }}\delta \left( {k_x  - \frac{\omega }{v}} \right)\left[ { -  \frac{{k_y }}{{k_{1z} }}{\bf{u}}_{\rm S}  + \frac{{k_{1z} }}{{{\beta k_0 } \varepsilon _1 }}\left( {{\bf{u}}_{\rm P}  - {\mathop{\rm sign}} \left( {z + d} \right)\frac{{k_\parallel  }}{{k_{1z} }}{\bf{e}}_z } \right)} \right].
\end{equation}
Note that, due to the delta function, the longitudinal wave number is
\begin{equation} 
k_{1z}  =  i \sqrt { {\frac{\omega^2 }{{v^2\gamma^2 }}} + k_y^2 }
\end{equation}
so that the field spectrum is entirely evanescent and  $|z+d|$ assures that the field fades both for $z<-d$ and $z>-d$.

\begin{figure*} \label{FigS4}
\includegraphics[width=0.5\textwidth]{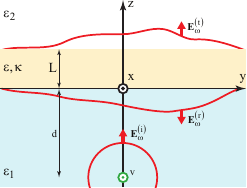}
\caption{MOA triggered by a fast electron interacting with a chiral slab in aloof configuration. The electron provides an incident MSF triggering the reflection and transmission of IPFs by the slab.}
\end{figure*}

Suppose now that the electron travels in the substrate of the setup reported in Fig.S4 in the presence of the chiral slab. The field in Eqs.(\ref{Ee}) satisfies Maxwell's equations in $z<0$ and, along the stripe $-d<z<0$, its Fourier representation of Eq.(\ref{FourierE}) turns into an angular spectrum representation 
\footnote{
Equation (\ref{FourierE}) is not an angular spectrum representation due to presence of the modulus and sign of $z+d$.  This is a consequence of the fact that the electron field is not source-free since it satisfies Maxwell's equations with the charge and current densities, $\rho _\omega ^{\left( \rm e \right)}$ and ${\bf{J}}_\omega ^{\left( \rm e \right)}$, associated to the moving electron, i.e
\[
\left\{ \begin{array}{l}
 \nabla  \cdot {\bf{E}}_\omega ^{\left(\rm e \right)}  = \frac{1}{{\varepsilon _0 \varepsilon _1 }}\rho _\omega ^{\left(\rm e \right)}  \\ 
 \nabla  \cdot {\bf{H}}_\omega ^{\left(\rm e \right)}  = 0 \\ 
 \nabla  \times {\bf{E}}_\omega ^{\left(\rm e \right)}  = i\omega \mu _0 {\bf{H}}_\omega ^{\left(\rm e \right)}  \\ 
 \nabla  \times {\bf{H}}_\omega ^{\left(\rm e \right)}  =  - i\omega \varepsilon _0 \varepsilon _1 {\bf{E}}_\omega ^{\left(\rm e \right)}  + {\bf{J}}_\omega ^{\left(\rm e \right)}  \\ 
 \end{array} \right. , \quad \quad
\left\{ \begin{array}{l}
 \rho _\omega ^{\left( \rm e \right)}  =  - \frac{e}{{2\pi \left| v \right|}}e^{i\frac{\omega }{v}x} \delta \left( y \right)\delta \left( {z + d} \right) \\ 
 {\bf{J}}_\omega ^{\left( \rm e \right)}  =  - \frac{e}{{2\pi }}e^{i\frac{\omega }{v}x} \delta \left( y \right)\delta \left( {z + d} \right){\bf{e}}_x  \\ 
 \end{array} \right. .
\]
However, in both the regions $z>-d$ and $z<-d$, $\rho _\omega ^{\left( e \right)}$ and ${\bf{J}}_\omega ^{\left( e \right)}$ vanish and the field locally admits an angular spectrum representation which depend on the region.
}
whose structure coincides with that of Eq.(\ref{i-field}) (since $z+d >0$). As a consequence the electron is the source of a field impinging onto the chiral slab whose spectral amplitudes are
\begin{eqnarray} \label{Uee}
 U_{\rm S}^{\left( \rm ei \right)}  &=& E_{\omega 0} \frac{{e^{ik_{1z} d} }}{{2k_\parallel  }}\left( { - \frac{{k_y }}{{k_{1z} }}} \right)\delta \left( {k_x  - \frac{\omega }{v}} \right), \nonumber \\ 
 U_{\rm P}^{\left( \rm ei \right)}  &=& E_{\omega 0} \frac{{e^{ik_{1z} d} }}{{2k_\parallel  }}\left( {\frac{{k_{1z} }}{{ {\beta k_0 } \varepsilon _1 }}} \right)\delta \left( {k_x  - \frac{\omega }{v}} \right),
\end{eqnarray}
which manifestly satisfy Eqs.(\ref{UMSF}), as expected since the electron field is a MSF. Therefore the fast electron triggers MOA whose efficiency is magnified by the evanescent character of the impinging field in the sub-Cherenkov regime $\beta < \frac{1}{\sqrt{\varepsilon_1}}$. The reflected field ${\bf{E}}_\omega ^{\left( \rm er \right)}$ is a IPF and its symmetric and antisymmetric parts can be evaluated by inserting Eqs.(\ref{Uee}) into Eqs.(\ref{MOAr}) thus getting
\begin{eqnarray} \label{SPOA}
 {\bf{E}}_\omega ^{\left( \rm e r \right){\rm S}}  &=& e^{i\frac{\omega }{v}x} \int\limits_{ - \infty }^{ + \infty } {dk_y } e^{ik_y y} \left\{ {e^{ik_{1z} z} \left[ {\tilde U_{\rm S}^{\left( \rm r \right){\rm A}} {\bf{u}}_{\rm S}  + \tilde U_{\rm P}^{\left( \rm r \right){\rm S}} \left( {{\bf{u}}_{\rm P}  + \frac{{k_\parallel  }}{{k_{1z} }}{\bf{e}}_z } \right)} \right]} \right\}_{k_x  = \frac{\omega }{v}},\nonumber  \\ 
 {\bf{E}}_\omega ^{\left( \rm e r \right){\rm A}}  &=& e^{i\frac{\omega }{v}x} \int\limits_{ - \infty }^{ + \infty } {dk_y } e^{ik_y y} \left\{ {e^{ik_{1z} z} \left[ {\tilde U_{\rm S}^{\left( \rm r \right){\rm S}} {\bf{u}}_{\rm S}  + \tilde U_{\rm P}^{\left( \rm r \right){\rm A}} \left( {{\bf{u}}_{\rm P}  + \frac{{k_\parallel  }}{{k_{1z} }}{\bf{e}}_z } \right)} \right]} \right\}_{k_x  = \frac{\omega }{v}}
\end{eqnarray}
where 
\begin{equation} \label{SAref}
\begin{array}{ll}
\displaystyle   {\tilde U_{\rm S}^{\left(\rm  r \right){\rm A}}  = E_{\omega 0} \left[ {\frac{{e^{ik_{1z} d} }}{{2k_\parallel  }}\left( { - \frac{{k_y }}{{k_{1z} }}R_{\rm SS} } \right)} \right]_{k_x  = \frac{\omega }{v}} ,} & 
\displaystyle   {\tilde U_{\rm P}^{\left(\rm  r \right){\rm S}}  = E_{\omega 0} \left[ {\frac{{e^{ik_{1z} d} }}{{2k_\parallel  }}\left( {\frac{{k_{1z} }}{{  {\beta k_0 }  \varepsilon _1 }}R_{\rm PP} } \right)} \right]_{k_x  = \frac{\omega }{v}} ,}  \\
\displaystyle    {\tilde U_{\rm S}^{\left(\rm  r \right){\rm S}}  = E_{\omega 0} \left[ {\frac{{e^{ik_{1z} d} }}{{2k_\parallel  }}\left( {n\frac{{k_{1z} }}{{  {\beta k_0 }  \varepsilon _1 }}R_{\rm SP} } \right)} \right]_{k_x  = \frac{\omega }{v}} ,} & 
\displaystyle    {\tilde U_{\rm P}^{\left(\rm  r \right){\rm A}}  = E_{\omega 0} \left[ {\frac{{e^{ik_{1z} d} }}{{2k_\parallel  }}\left( { - n\frac{{k_y }}{{k_{1z} }}R_{\rm PS} } \right)} \right]_{k_x  = \frac{\omega }{v}} ,}  \\
\end{array}
\end{equation}
are the symmetric and antisymmetric parts of the S and P amplitudes of the reflected field.
\begin{figure*} \label{FigS5}
\includegraphics[width=1\textwidth]{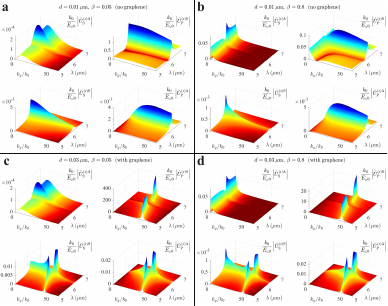}
\caption{Symmetric and antisymmetric parts of the S and P amplitudes of the IPF reflected by a chiral slab interacting with an aloof electron. The normalized amplitudes are plotted as functions of the normalized wavenumber $k_y/k_0$ (for $k_y>0$) and the wavelength $\lambda$, for different parameters $d,\beta$ of the electron interacting with the same substrate-slab-superstrate (with and without graphene) considered in Fig.S2.}
\end{figure*}

In Fig.S5 we plot the four spectral amplitudes of Eqs.(\ref{SAref}), as functions of the normalized wavenumber $k_y/k_0$ (for $k_y>0$) and the wavelength $\lambda$, for different parameters $d,\beta$ of the electron interacting with the same substrate-slab-superstrate (with and without graphene) considered in Fig.S2. In panels (a) and (b) graphene is absent and the electron trajectory is located at $d=0.01 \: \mu{\rm m}$ away from the chiral slab interface with $\beta = 0.05$ (a) and $\beta = 0.8$ (b). In the first case of lower electron energy (a) the amplitude $\tilde U_{\rm P}^{\left(\rm r \right){\rm S}}$ is dominant and hence the antisymmetric part ${\bf{E}}_\omega ^{\left( \rm r \right){\rm A}}$ is small compared to ${\bf{E}}_\omega ^{\left( \rm r \right){\rm S}}$ so that MOA efficiency is expected to be low. On the other hand, in the second case of higher electron energy (b) $\tilde U_{\rm P}^{\left(\rm r \right){\rm S}}$ is considerably smaller than the previous case and it is comparable with $\tilde U_{\rm P}^{\left(\rm r \right){\rm A}}$ (while $\tilde U_{\rm S}^{\left(\rm r \right){\rm A}}$ and $\tilde U_{\rm S}^{\left(\rm r \right){\rm S}}$ are comparable) so that MOA efficiency is higher than case (a). We conclude that MOA efficiency increases by increasing the electron energy and this agrees with the discussion of subsection S3.3 since the S/P amplitudes ratio of the electron (incident) field
(see Eqs.(\ref{Uee}))
\begin{equation}
\frac{{\left| {U_{\rm S}^{\left(\rm ei \right)} } \right|}}{{\left| {U_{\rm P}^{\left(\rm ei \right)} } \right|}} = \frac{{\beta \varepsilon _1 }}{{\left( {\frac{1}{{\beta ^2 }} - \varepsilon _1 } \right)\frac{{k_0 }}{{\left| {k_y } \right|}} + \frac{{\left| {k_y } \right|}}{{k_0 }}}}
\end{equation}
is an increasing function of $\beta$. In panels (c) and (d) of Fig.S5 graphene is present and the electron trajectory is located at $d=0.03 \: \mu{\rm m}$ away from the chiral slab interface with $\beta = 0.05$ (c) and $\beta = 0.8$ (d). The same reasoning as in the graphene-free situations of panels (a) and (b) leads to the conclusion that the MOA efficiency is higher at larger electron energy and it also benefits of the graphene plasmon resonance.

The setup encompassing a graphene sheet is however even more interesting since the combination of graphene plasmon resonance and field dissymmetry triggers a novel near field interferometric phenomenon which can further increase MOA efficiency. Panels (c) and (d) of Fig.S5 manifestly signal the graphene plasmon resonance through a peak centered at a specific parallel wavenumber $k_y=k_p (\lambda)$. Now the resonance corresponds to the minimum of the determinant of the matrix  $\underline{\underline M}$ in the first of Eqs.(\ref{equa_matr}) which is invariant under the change ${\bf k}_\parallel \rightarrow - \mathcal{R} {\bf k}_\parallel$ so that  $k_y=k_p$ and $k_y=-k_p$ equivalently contribute to the graphene plasmon resonance. If we neglect for simplicity the width of the resonance peaks and we retain in Eqs.(\ref{SPOA}) only the contributions from the wavenumbers $k_y=k_p$ and $k_y=-k_p$ we get
\begin{eqnarray}
 {\bf{E}}_\omega ^{\left(\rm r \right){\rm S}}  &\cong & \Delta k\left\{ {e^{ik_x x} e^{ik_{1z} z} \left[ {\tilde U_{\rm S}^{\left(\rm r \right){\rm A}} {\bf{u}}_{\rm S}  + \tilde U_{\rm P}^{\left(\rm r \right){\rm S}} \left( {{\bf{u}}_{\rm P}  + \frac{{k_\parallel  }}{{k_{1z} }}{\bf{e}}_z } \right)} \right]} \right\}_{{\bf{k}}_\parallel   = \frac{\omega }{v}{\bf{e}}_x  + k_p {\bf{e}}_y } 2 \: {\rm cos} \left( {k_p y} \right), \nonumber \\ 
 {\bf{E}}_\omega ^{\left(\rm r \right){\rm A}}  &\cong & \Delta k\left\{ {e^{ik_x x} e^{ik_{1z} z} \left[ {\tilde U_{\rm S}^{\left(\rm r \right){\rm S}} {\bf{u}}_{\rm S}  + \tilde U_{\rm P}^{\left(\rm r \right){\rm A}} \left( {{\bf{u}}_{\rm P}  + \frac{{k_\parallel  }}{{k_{1z} }}{\bf{ e}}_z } \right)} \right]} \right\}_{{\bf{k}}_\parallel   = \frac{\omega }{v}{\bf{e}}_x  + k_p {\bf{e}}_y } 2i \: {\rm sin} \left( {k_p y} \right), 
\end{eqnarray}
where $\Delta k$ is the resonance width and the symmetry and antisymmetry of the integrands of ${\bf{E}}_\omega ^{\left(\rm r \right){\rm S}}$ and ${\bf{E}}_\omega ^{\left(\rm r \right){\rm A}}$ have respectively been exploited (see Eqs.(\ref{uREF}) and (\ref{SAref})). We conclude that the interference between the counterpropagating graphene plasmon polaritons of parallel wavenumbers $k_y=k_p$ and $k_y=-k_p$ endows the symmetric and the antisymmetric parts of the reflected field with cosinusoidal and sinusoidal profiles, respectively, along the $y-$ axis whose spatial period is $2 \pi / k_p$. What is remarkable here is that at the points where $|{\bf{E}}_\omega ^{\left(\rm r \right){\rm A}}|$ is maximum, $|{\bf{E}}_\omega ^{\left(\rm r \right){\rm S}}|$ is minimum and consequently the dissymmetry factor $\Upsilon$ of Eq.(\ref{dissF}) benefits of an additional local enhancement thus raising MOA efficiency.

\section{MOA detection through cathodoluminescence}
MOA accompanying the excitation of a chiral slab by means of a fast electron is intrinsically a near field phenomenon (as discussed above) and hence its detection technique can be borrowed from the number of currently available nanophotonics strategies. In this work we consider the standard nanoantenna techique where a nanoparticle in the substrate experiences the near field and it outcouples radiation in turn detectable in the superstrate as cathodoluminescence. By placing the nanoparticle at two symmetrical positions alongside the electron trajectory, two unequal chatodoluminescence signals are collected, their difference amounting to MOA detection.

We consider a plasmonic nanoparticle of radius $a$ lying upon the graphene sheet with its center located at ${\bf{r}}_{\rm  NP}  = h{\bf{e}}_y  - a{\bf{e}}_z$ in the substrate, as sketched in Fig.S6. The nanoparticle dielectric permittivity is described by the Drude model $\varepsilon _{\rm NP} \left( \omega  \right) = 1 - \frac{{\omega _p^2 }}{{\omega ^2  + i\omega \Gamma }}$ which accurately describes transparent conductors with plasma frequency $\omega _p$ in the mid-infrared. Since the radius $a$ is much smaller than the mid-infrared wavelengths and the near field will not self-consistently display large field gradients, we here resort to the electrostatic (noretarded) approximation where the nanoparticle is modelled by a point dipole located at ${\bf{r}}_{\rm  NP}$ whose dipole moment (in the frequency domain) is ${\bf{p}}_\omega  = \alpha {\bf{E}}_\omega ^{\left( {\rm ext} \right)}$ where ${\bf{E}}_\omega ^{\left( {\rm ext} \right)}$ is the field experienced by the dipole (without self-field) and $\alpha  = 4\pi \varepsilon _0 \varepsilon _1 a^3 \left( {\frac{{\varepsilon _{\rm NP}  - \varepsilon _1 }}{{\varepsilon _{\rm NP}  + 2\varepsilon _1 }}} \right)$ is the well-known polarizability of the sphere. The field produced by the dipole in the substrate ($z<0$) is
\begin{equation}
{\bf{E}}_\omega ^{\left( \rm p \right)}  = \left( {k_0^2 \varepsilon _1  + \nabla \nabla  \cdot } \right)\left( {\frac{1}{{4\pi \varepsilon _0 \varepsilon _1 }}\frac{{e^{ik_0 \sqrt {\varepsilon _1 } \left| {{\bf{r}} - {\bf{r}}_{\rm NP} } \right|} }}{{\left| {{\bf{r}} - {\bf{r}}_{\rm NP} } \right|}}{\bf{p}}_\omega  } \right)
\end{equation}
which, after using the Weyl representation of the spherical wave

\begin{equation}
\frac{{e^{ik_0 \sqrt{\varepsilon_1} r} }}{r} = \frac{i}{{2\pi }}\int {d^2 {\bf{k}}_ \parallel  e^{i{\bf{k}}_ \parallel   \cdot {\bf{r}}_ \parallel } } \frac{{e^{ik_{1z} \left| z \right|} }}{{k_{1z} }},
\end{equation}
can be casted as
\begin{eqnarray} \label{dipoleF}
{\bf{E}}_\omega ^{\left( \rm p \right)}  &=& \int {d^2 {\bf{k}}_\parallel  } e^{i{\bf{k}}_\parallel \cdot {\bf{r}}_\parallel } e^{ik_z \left| {z + a} \right|} \frac{{ik_0 e^{ - ik_y h} }}{{8\pi ^2 \varepsilon _0 \varepsilon _1 }} \left[ \left( {\frac{{k_0 \varepsilon _1 }}{{k_{1z} }}{\bf{u}}_{\rm S} } \right) \cdot {\bf{p}}_\omega  {\bf{u}}_{\rm S} \right. + \\
  &&+  \left. \left( {\frac{{k_{1z} }}{{k_0 }}{\bf{u}}_{\rm P}  - \frac{{k_\parallel  }}{{k_0 }}{\mathop{\rm sgn}} \left( {z + a} \right){\bf{e}}_z } \right) \cdot {\bf{p}}_\omega  \left( {\bf{u}}_{\rm P}  - \frac{{k_\parallel  }}{{k_{1z} }}{\mathop{\rm sgn}} \left( {z + a} \right){\bf{e}}_z  \right) \right].
\end{eqnarray}
In analogy with the electron field of Eq.(\ref{FourierE}), 
\begin{figure*} \label{FigS6}
\includegraphics[width=0.7\textwidth]{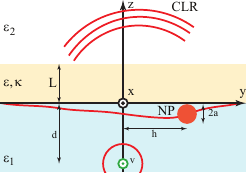}
\caption{Cathodoluminesce radiation (CLR) produced by a nanoparticle (NP) in the substrate.}
\end{figure*}
the Fourier representation of the dipole field in Eq.(\ref{dipoleF}) reduces to an angular spectrum representation of the form of Eq.(\ref{i-field}) along the stripe $-a < z <0$ 
from which its S and P amplitudes are 
\begin{eqnarray} \label{Upi}
 U_{\rm S}^{\left( \rm pi \right)}  &=& \frac{{ik_0 }}{{8\pi ^2 \varepsilon _0 \varepsilon _1 }}e^{ - ik_y h} e^{ik_{1z} a} \left( {\frac{{k_0 \varepsilon _1 }}{{k_{1z} }}{\bf{u}}_{\rm S} } \right) \cdot {\bf{p}}_\omega, \nonumber    \\ 
 U_{\rm P}^{\left( \rm pi \right)}  &=& \frac{{ik_0 }}{{8\pi ^2 \varepsilon _0 \varepsilon _1 }}e^{ - ik_y h} e^{ik_{1z} a} \left( {\frac{{k_{1z} }}{{k_0 }}{\bf{u}}_{\rm P}  - \frac{{k_\parallel  }}{{k_0 }}{\bf{e}}_z } \right) \cdot {\bf{p}}_\omega   
\end{eqnarray}
Inserting Eqs.(\ref{Upi}) into Eqs.(\ref{react}), we obtain the reflected (pr) and transmitted (pt) dipole fields 
\begin{eqnarray} \label{Eprt}
 {\bf{E}}_\omega ^{\left(\rm {pr} \right)}  &=& G^{\left(\rm {pr} \right)} {\bf{p}}_\omega  \quad \left( {z < 0} \right), \nonumber \\ 
 {\bf{E}}_\omega ^{\left(\rm {pt} \right)}  &=& G^{\left( \rm {pt} \right)} {\bf{p}}_\omega  \quad \left( {z > L} \right), 
\end{eqnarray}
where, using the dyadic notation $( {{\bf{ab}} } ){\bf{c}} = \left( {{\bf{b}} \cdot {\bf{c}}} \right) {\bf{a}}$,
\begin{eqnarray} \label{Grt}
 G^{\left(\rm {pr} \right)}  &=& \frac{{ik_0 }}{{8\pi ^2 \varepsilon _0 \varepsilon _1 }}\int {d^2 {\bf{k}}_\parallel  } e^{i{\bf{k}}_\parallel   \cdot {\bf{r}}_\parallel  } e^{ - ik_{1z} z} e^{ - ik_y h} e^{ik_{1z} a} \left\{ \left[ {R_{\rm SS} {\bf{u}}_{\rm S}  + nR_{\rm PS} \left( {{\bf{u}}_{\rm P}  + \frac{{k_\parallel  }}{{k_{1z} }}{\bf{e}}_z } \right)} \right]\left( {\frac{{k_0 \varepsilon _1 }}{{k_{1z} }}{\bf{u}}_{\rm S} } \right) + \right. \nonumber \\
&&\left. + \left[ {nR_{\rm SP} {\bf{u}}_{\rm S}  + R_{\rm PP} \left( {{\bf{u}}_{\rm P}  + \frac{{k_\parallel  }}{{k_{1z} }}{\bf{ e}}_z } \right)} \right]\left( {\frac{{k_{1z} }}{{k_0 }}{\bf{u}}_{\rm P}  - \frac{{k_\parallel  }}{{k_0 }}{\bf{e}}_z } \right) \right\}, \nonumber \\ 
 G^{\left(\rm {pt} \right)}  &=& \frac{{ik_0 }}{{8\pi ^2 \varepsilon _0 \varepsilon _1 }}\int {d^2 {\bf{k}}_\parallel  } e^{i{\bf{k}}_\parallel   \cdot {\bf{r}}_\parallel  } e^{ik_{2z} \left( {z - L} \right)} e^{ - ik_y h} e^{ik_{1z} a} \left\{ \left[ {T_{\rm SS} {\bf{u}}_{\rm S}  + nT_{\rm PS} \left( {{\bf{u}}_{\rm P}  - \frac{{k_\parallel  }}{{k_{2z} }}{\bf{e}}_z } \right)} \right]\left( {\frac{{k_0 \varepsilon _1 }}{{k_{1z} }}{\bf{u}}_{\rm S} } \right)  + \right. \nonumber \\
&&\left.  + \left[ {nT_{\rm SP} {\bf{u}}_{\rm S}  + T_{\rm PP} \left( {{\bf{u}}_{\rm P}  - \frac{{k_\parallel  }}{{k_{2z} }}{\bf{ e}}_z } \right)} \right]\left( {\frac{{k_{1z} }}{{k_0 }}{\bf{u}}_{\rm P}  - \frac{{k_\parallel  }}{{k_0 }}{\bf{e}}_z } \right)  \right\}.
\end{eqnarray}
The dyadics $G^{\left(\rm {pr} \right)}$ and $G^{\left(\rm {pt} \right)}$ fully describe the electromagnetic response of the chiral slab in the achiral environment and they may be easily evaluated numerically in the near field. In addition, their Fourier representations are directly suitable to the far field analysis (see below). In order to evaluate the dipole moment ${\bf p}_{\omega}$ induced by the fast electron and the chiral slab, note that the nanoparticle experiences the external field
\begin{equation}
{\bf{E}}_\omega ^{\left( {\rm ext} \right)}  = \left[ {{\bf{E}}_\omega ^{\left(\rm e \right)}  + {\bf{E}}_\omega ^{\left( {\rm er} \right)}  + {\bf{E}}_\omega ^{\left( {\rm pr} \right)} } \right]_{{\bf{r}} = {\bf{r}}_{\rm NP} } 
\end{equation}
so that, using the nanoparticle polarizability $\alpha$ and the first of Eqs.(\ref{Eprt}), we get
\begin{equation}
{\bf{p}}_\omega   = \left\{ {\left[ {\frac{1}{\alpha }I - G^{\left( {\rm pr} \right)} } \right]^{ - 1} \left( {{\bf{E}}_\omega ^{\left(\rm e \right)}  + {\bf{E}}_\omega ^{\left( {\rm er} \right)} } \right)} \right\}_{{\bf{r}} = {\bf{r}}_{\rm NP} },
\end{equation}
where $I$ is the identity dyadic. Such equation reveals that the nanoparticle provides an ideal tool for probing MOA since its dipole moment linearly depend on the total electron field ${\bf{E}}_\omega ^{\left(\rm e \right)}  + {\bf{E}}_\omega ^{\left( {\rm er} \right)}$ evaluated at the center of the nanoparticle.

Since we are considering the sub-Cherenkov regime and the aloof electron does not produce transition radiation, cathodoluminesce radiation is only due to the nanoparticle. Radiation detection is usually performed in the superstrate and consequently we focus on ${\bf{E}}_\omega ^{\left(\rm {pt} \right)}$. After introducing polar coordinates $x = r\sin \theta \cos \varphi$, $y = r\sin \theta \sin \varphi$, $z = r\cos \theta$ together with their coordinate unit vectors ${\bf{e}}_r ,{\bf{e}}_\theta  ,{\bf{e}}_\varphi$ and using the relation
\begin{equation}
\int {d^2 {\bf{k}}_\parallel  } e^{i{\bf{k}}_\parallel   \cdot {\bf{r}}_\parallel  } e^{ik_{2z} z} F\left( {{\bf{k}}_\parallel  } \right) = \frac{{e^{ik_0 \sqrt {\varepsilon _2 } r} }}{r}\left( {\frac{{2\pi }}{i}} \right)F\left( {k_0 \sqrt {\varepsilon _2 } \sin \theta \frac{{{\bf{r}}_\parallel  }}{{r_\parallel  }}} \right)k_0 \sqrt {\varepsilon _2 } \cos \theta 
\end{equation}
providing the leading order of the  asymptotic expansion for $k_0 r \rightarrow \infty$ and $\cos \theta >0$ of the Fourier integral,  from the seconds of Eqs.(\ref{Eprt}) and (\ref{Grt}), ${\bf{E}}_\omega ^{\left(\rm {pt} \right)}$ gets the usual far field expression
\begin{equation} \label{farf}
{\bf{E}}_\omega ^{\left( {\rm pt} \right)}  = \frac{{e^{ik_0 \sqrt {\varepsilon _2 } r} }}{{k_0 r}}{\bf{f}}\left( {\theta ,\varphi } \right)
\end{equation}
where
\begin{eqnarray}
 {\bf{f}} &=& \left[ { \cos \theta \left( {T_{\rm SS}  E_{\rm S}  + nT_{\rm SP} E_{\rm P} } \right) {\bf{e}}_\varphi   + \left( {nT_{\rm PS} E_{\rm S}  + T_{\rm PP} E_{\rm P} } \right){\bf{e}}_\theta  } \right]_{k_\parallel   = k_0 \sqrt {\varepsilon _2 } \sin \theta }, \nonumber  \\ 
 \left( {\begin{array}{@{\mkern0mu} c @{\mkern0mu}}
   {E_{\rm S} }  \\
   {E_{\rm P} }  \\
\end{array}} \right) &=& \frac{{k_0^3 \sqrt {\varepsilon _2 } }}{{4\pi \varepsilon _0 \sqrt {\varepsilon _1 } }}
e^{ - ik_0 \left[ {\sqrt {\varepsilon _2 } L\cos \theta  + \sqrt {\varepsilon _2 } h\sin \theta \sin \varphi  - a\sqrt {\varepsilon _1  - \varepsilon _2 \sin ^2 \theta } } \right]} 
  \left( {\begin{array}{@{\mkern0mu} c @{\mkern0mu}}
\displaystyle   {\frac{{ - \sin \varphi p_{\omega x}  + \cos \varphi p_{\omega y} }}{{\sqrt {1 - \frac{{\varepsilon _2 }}{{\varepsilon _1 }}\sin ^2 \theta } }}}  \\
\displaystyle   {\sqrt {1 - \frac{{\varepsilon _2 }}{{\varepsilon _1 }}\sin ^2 \theta } \left( {\cos \varphi p_{\omega x}  + \sin \varphi p_{\omega y} } \right) - \sqrt {\frac{{\varepsilon _2 }}{{\varepsilon _1 }}} \sin \theta p_{\omega y} } 
\end{array}} \right). \nonumber \\ 
\end{eqnarray}
The total energy emitted by cathodoluminescence per incoming electron is 
\begin{equation}
U = \int\limits_{ - \infty }^\infty  {dt} \int {d\Omega } \: r^2 {\bf{ e}}_r  \cdot \left[ {{\bf{E}}\left( {{\bf{r}},t} \right) \times {\bf{H}}\left( {{\bf{r}},t} \right)} \right] = 
\int\limits_0^\infty  {d\lambda } \left( {\frac{{hc}}{\lambda }} \right)\int {d\Omega } \;\frac{{dN}}{{d\Omega  d\lambda }}
\end{equation}
where $\frac{hc}{\lambda}$ is the photon energy quantum and $\frac{{dN}}{{ d\Omega d\lambda}} = \frac{{4\pi }}{{\hbar \lambda }} r^2 {\bf{ e}}_r  \cdot {\mathop{\rm Re}\nolimits} \left( {{\bf{E}}_\omega   \times {\bf{H}}_\omega ^* } \right)$ is the number of photons emitted per incoming electron, per per unit of solid angle of emission and per unit of photon wavelength. By using Eq.(\ref{farf}) and the far field relation ${\bf{H}}_\omega^{\rm (pt)}   = \frac{{\sqrt \varepsilon_2  }}{{Z_0 }}{\bf{ e}}_r  \times {\bf{E}}_\omega^{\rm (pt)} $, we get in the superstrate $\frac{{dN}}{{d\Omega d\lambda }} = \frac{{\lambda \sqrt \varepsilon_2  }}{{\pi \hbar Z_0 }}\left| {\bf f } \right|^2$  which, after integration over the solid angle $\cos \theta >0$, yields
\begin{equation}
\Gamma  = \frac{{\lambda \sqrt {\varepsilon _2 } }}{{\pi \hbar Z_0 }}\int\limits_{\cos \theta  > 0} {d\Omega } \left| {\bf{f}} \right|^2,
\end{equation}
i.e. the number of photons emitted per incoming electron in the superstrate per unit photon wavelength .

Suppose now that two symmetrical situations are considered where the same nanoparticle is placed at ${\bf{r}}_{\rm  NP}^{(+)}  = h{\bf{e}}_y  - a{\bf{e}}_z$ and at ${\bf{r}}_{\rm  NP}^{(-)}  = -h{\bf{e}}_y  - a{\bf{e}}_z$. Due to MOA the corresponding nanoparticle dipole moments ${\bf p}_\omega^{(+)}$ and ${\bf p}_\omega^{(-)}$ are necessarily different and consequently the associated cathodoluminescence signals $\Gamma(h)$ and $\Gamma(-h)$ are different as well. Therefore the dissymmetry factor
\begin{equation}
\delta \Gamma \left( h \right) =  - 2\frac{{\Gamma \left( {  h} \right) - \Gamma \left( -h \right)}}{ {\Gamma \left( {  h} \right) + \Gamma \left( - h \right)}}
\end{equation}
effectively provides a quantitative measurement of MOA efficiency and chiral sensing.

\section{Infrared graphene local optical response}
At infrared frequencies the optical response of graphene is dominated by the conical band structure ${\cal E} = \pm v_{\rm F} |{\bf p}_\parallel|$ around the two Dirac points of the first Brillouin zone, where $v_{\rm F} \approx 9 \cdot 10^5$ m$/$s is the Fermi velocity and ${\cal E},{\bf p}_\parallel$ are the electron energy and momentum, respectively. In the present investigation we mainly focus on the infrared range of wavelengths $5 \: \mu m < \lambda < 20 \: \mu m$. While in undoped graphene the Fermi energy lies at the Dirac points, injection of charge carriers through electrical gating \cite{Chen2011} or chemical doping \cite{Liu2011} efficiently shifts the Fermi level up to $E_{\rm F} \approx 1$ eV owing to the conical dispersion and the 2D electron confinement. The response of graphene to photons of energy $\hbar \omega$ and in-plane momentum $\hbar {\bf k}_{\parallel}$ is described by the surface conductivity $\sigma_{\rm G}({\bf k}_{\parallel},\omega)$  which is generally affected both by intraband and interband electron dynamics. The dependence of $\sigma_{\rm G}$ on ${\bf k}_{\parallel}$ physically arises from electron-hole pairs excitation and it generally yields unwanted absorption (Landau damping) and nonlocal effects. However, if 
\begin{equation} \label{locality}
\frac{{ {k_ \parallel  } }}{{k_{\rm F} }} < \frac{{\hbar \omega }}{{E_{\rm F} }} < 2 - \frac{{ {k_ \parallel  } }}{{k_{\rm F} }}
\end{equation}
where $k_{\rm F}  = E_{\rm F} /\hbar v_{\rm F}$ is the Fermi wave number, the photon momentum is too small to trigger intraband transitions and interband transitions are forbidden by the Pauli exlusion principle \cite{Hwang}. Once interacting with photons satisfying Eq.(\ref{locality}), nonlocal effects can be neglected and graphene displays a marked metal-like behavior with long relaxation time $\tau = \mu E_{\rm F}/ev_{\rm F}^2$, where $\mu$ is the electron mobility, which conversely to noble metals can reach the picosecond time scale at moderate doping and purity (affecting electron mobility) \cite{JavierACSPhot}. In such local regime, random phase approximation (RPA) provides for the graphene conductivity the integral expression
\begin{equation} \label{sigma}
\sigma_{\rm G} (\omega) = \frac{-ie^2}{\pi\hbar^2(\omega+i/\tau)}\int_{-\infty}^{+\infty}d{\cal E} \left\{|{\cal E}| \frac{ \partial f_{\cal E} }{\partial {\cal E}} +    \frac{ {\rm sign} \left(\cal E\right) }{1 - 4 {\cal E}^2/[\hbar(\omega+i/\tau)]^2} f_{\cal E} \right\},
\end{equation}
where $f_{\cal E} = \{ {\rm exp}[({\cal E} - E_{\rm F})/k_{\rm B} T] +1 \}^{-1}$ is the Fermi function ($k_{\rm B}$ is the Boltzmann constant and $T$ is the temperature). In our analysis, we focus on photon-graphene interactions satysfing Eq.(\ref{locality}) and accordingly we model graphene surface conductivity by means of Eq.(\ref{sigma}).

It is worth noting that, in the regime where the Fermi energy is greater than the photon energy ($E_{\rm F}  > \hbar \omega$), Eq.(\ref{locality}) can be casted as 
\begin{equation} \label{locality2}
k_\parallel   < \left( {\frac{c}{{v_F }}} \right)k_0  \simeq 333\,k_0,
\end{equation}
which specifies the  wavevector range of those photons not triggering  nonlocal effects at frequency $\omega$.